\documentclass[aip,pof,amsmath,amssymb,preprint]{revtex4-1}
\usepackage{epsfig}
\usepackage{color}
\usepackage{graphicx}
\usepackage{subfigure}
\usepackage{dcolumn}
\usepackage{bm}
\def\om {\omega}

\begin{document}

\title{Fluid flows in a librating cylinder}

\author{Alban Sauret} \email[Electronic mail: ]{sauret@irphe.univ-mrs.fr}
\author{David C\'ebron}
\author{Michael Le Bars}
\author{St\'ephane Le Diz\`es}
\affiliation{Institut de Recherche sur les Ph\'enom\`enes Hors \'Equilibre,  UMR 7342,
CNRS and \break
Aix-Marseille Univ., 49, rue F. Joliot-Curie, F-13013 Marseille, France}

\date{\today}
\begin{abstract}
The flow in a cylinder driven by time harmonic oscillations of the rotation rate, called longitudinal librations, is investigated. Using a theoretical approach and axisymmetric numerical simulations, we study two distinct phenomena appearing in this librating flow. First, we investigate the occurrence of a centrifugal instability near the oscillating boundary, leading to the so-called Taylor-G\"ortler vortices. A viscous stability criterion is derived and  compared to numerical results obtained for various libration frequencies and Ekman numbers. The strongly nonlinear regime well above the instability threshold is also documented. {{We show that a new mechanism of spontaneous generation of inertial waves in the bulk could exist when the sidewall boundary layer becomes turbulent}}. Then, we analyse the librating flow below the instability threshold and characterize the mean zonal flow correction induced by the nonlinear interaction of the boundary layer flow with itself. In the frequency regime where inertial modes are not excited, we show that the mean flow correction in the bulk is a uniform rotation, independent of the Ekman number and cylinder aspect ratio, in perfect agreement with the analytical results of Wang  [J. Fluid. Mech., {41}, pp. 581 - 592, 1970]. When inertial modes are resonantly excited, the mean flow correction is found to have a more complex structure. Its amplitude still scales as the square of the libration amplitude but now depends on the Ekman number.
\end{abstract}

%\pacs{Valid PACS appear here}
\maketitle

\label{firstpage}

%%%%%%%%%%%%%%%%%%%%%%%%%%%%%%%%%%%%%%%%%%
%%%%%%%%%%%%        Introduction      %%% %%%%%%%%%%%%%%%%
%%%%%%%%%%%%%%%%%%%%%%%%%%%%%%%%%%%%%%%%%%
\section{Introduction}

Rotating flows generically support oscillatory motions called inertial waves\cite{greenspanbook,messio2008}, whose frequencies range between plus and minus twice the spin frequency. 
{{  
When the rotating flow is limited by boundaries, these waves can combine to form inertial modes whose structure and properties depend on the geometry of the container. These inertial modes
are well-characterized in simple geometries only, as the cylinder or the sphere\cite{greenspanbook}. In other geometries, as the spherical shell for instance, the singular structure that some of 
them exhibit is still the subject of active research \cite{rieutord1997,rieutord2001,rieutord2010}. 
}}
Usually damped by viscosity, inertial modes can nevertheless be excited by various harmonic forcings such as the natural processes of libration, precession and tides in planetary fluid cores, providing that their azimuthal period $m$ and temporal frequency $\omega$ are close to those of the forcing. Once a mode is resonantly excited, its nonlinear self-interaction generates an intense axisymmetric flow referred to as the zonal flow (see Morize et al.\cite{morize2010}). In the absence of 
{{ inertial mode resonance,
}} 
harmonic forcing can also generate a zonal flow through nonlinear effects in the Ekman layer\cite{busse2010,sauret2010,calkins2010,noir2010,wang1970,busse1968,suess1971,busse2010b}. The study of the zonal flow with and without resonance is of interest in a geophysical context, as it provides constraints on the possible dynamics of liquid cores, internal oceans and atmospheres of planets\cite{rambaux2011}. It is also important in the aeronautical context of rotating flying objects\cite{gans1984} since the internal fluid dynamics could influence their trajectory.

In this paper, we focus on a particular forcing: the time harmonic oscillations of the rotation rate of a container, called longitudinal librations.   
The early studies of librational forcing mainly focused on the excitation of inertial modes in a sphere or spherical shell. The first experimental study of Aldridge \& Toomre\cite{aldridge1969} confirmed the theoretical resonance of inertial modes in a sphere and evidenced the presence of Taylor-G\"ortler vortices near the outer boundary 
{{ with axes oriented in the azimuthal direction of the sphere\cite{aldridgephd}.
}}
These experimental results were then confirmed numerically by Rieutord\cite{rieutord1991} and extended by Tilgner\cite{tilgner1999} who studied the linear response to a librational forcing in a spherical shell and investigated the attractors associated with these inertial modes. More recently, Noir et al.\cite{noir2009} investigated experimentally by direct flow visualization, using Kalliroscope particles, the appearance of Taylor-G\"ortler vortices in a spherical container. This study was then complemented by the numerical work of Calkins et al.\cite{calkins2010} who also found a nontrivial zonal flow in the interior. By the same time, using methods developed for a precessing sphere \cite[][]{busse1968}, Busse\cite{busse2010} obtained an analytical solution of the flow in a librating sphere in the limit of small libration frequency, when inertial modes are not excited. This weakly non-linear theory was then confirmed by Sauret et al.\cite{sauret2010} using a combined numerical and experimental study. 

In addition to these studies in spherical geometries of direct geo/astrophysical interest, the case of a librating cylinder has also been investigated. In a little known paper, Wang\cite{wang1970} already demonstrated theoretically and experimentally that the time-harmonic forcing generates a zonal flow in the bulk. He showed that this flow comes from nonlinear interactions in the top and bottom Ekman layers of the cylinder. He also provided an estimate for the amplitude of the zonal flow in the interior for arbitrary values of the frequency of libration in the absence of inertial mode forcing. The limit of small libration frequency compared to the mean rotation rate was recently recalculated by Busse\cite{busse2010b}. Noir et al.\cite{noir2010} studied experimentally, via direct flow visualization and LDV measurements, {{the mean zonal flow}} at libration frequencies comparable to spin rate. 
{{ 
They also analysed  the generation of Taylor-G\"ortler vortices and  suggested a scaling law for the appearance of the vortices  different from that in the sphere. 
}}
More recently, Lopez \& Marques\cite{lopez2011} performed a three dimensional numerical study of a librating cylinder at moderate Ekman number ($E=10^{-4}$) for three values of libration frequency and various values of the libration amplitude. They observed the spontaneous generation of inertial modes for libration frequency above $\omega>2$ and proposed a mechanism for the excitation of inertial modes based on a period doubling process. The present paper comes as a complement to these works. 

The paper is organized as follows. In section 2, we introduce the governing equations, the notations and the numerical method. In section 3, we study the occurrence of Taylor-G\"ortler vortices at the outer boundary. We derive a local stability criterion which is systematically tested as a function of the libration amplitude, the libration frequency and the Ekman number. Then we compare this criterion to numerical results and available experimental data. The spontaneous generation of inertial waves induced by the Taylor-G\"ortler vortices is discussed in section 4. Section 5 is devoted to the analysis of the zonal flow  {{over a wide range of libration frequencies, $\omega \in[0.05;25]$}}. We show that in the absence of inertial mode resonance, the numerical results are well described by the analytical prediction of Wang\cite{wang1970} in the bulk. Configurations where inertial modes are present are also documented in this section.

%%%%%%%%%%%%%%%%%%%%%%%%%%%%%%%%%%%%%%%%%%%
%%%%%%%%%%%%%   Mathematical formulation %%%%%%%%%%%%%%%%
%%%%%%%%%%%%%%%%%%%%%%%%%%%%%%%%%%%%%%%%%%%

\section{Mathematical formulation and numerical model}

\subsection{Governing equations}

\begin{center}
\begin{figure}
\begin{center}\includegraphics{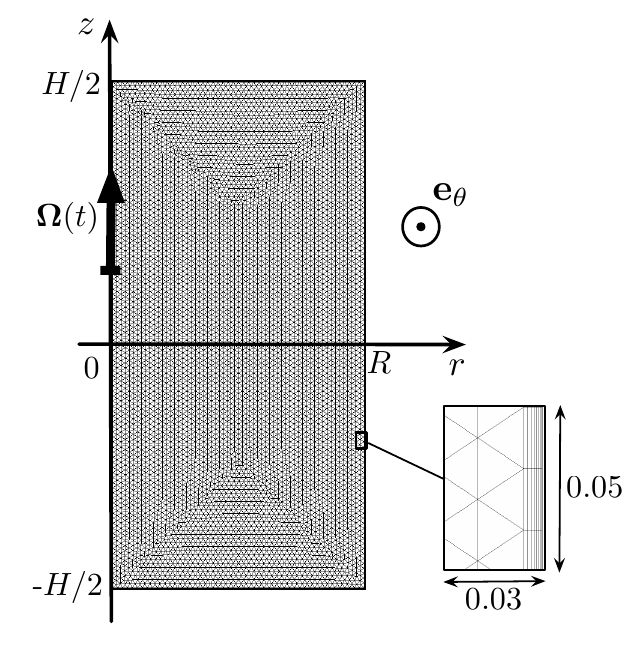}\end{center}
\caption{Schematic view of the cylindrical cavity and the cylindrico-polar coordinate ($r$,$\theta$,$z$). The $z$-axis is the symmetry axis {{and of rotation}}. The meshgrid used for the numerical simulations is also represented, with a denser mesh near the axis of rotation and near the outer boundaries (see section \ref{numericmodel} for details).}\label{schema_cylinder}
\end{figure}
\end{center}

We consider a homogeneous and incompressible fluid of kinematic viscosity $\nu$ in a cylinder of radius $R$ and height $H$ (see figure \ref{schema_cylinder}) rotating at the mean rotation rate $\Omega_0$  {{along its symmetry axis}}. Using $R$ and ${\Omega_0}^{-1}$ as the length scale and the time scale, respectively, the rotation rate of the cylinder is given by
\begin{eqnarray}
\label{vitesse_angulaire}
\boldsymbol{\Omega}(t)=\left[1+\epsilon\,\cos(\omega\,t)\right]\boldsymbol{e_z},
\end{eqnarray} 
\noindent where $\omega=\omega_{lib}/\Omega_0$ is the dimensionless libration frequency, $\epsilon=\Delta \theta\,\omega$ is the amplitude of the librational forcing with $\Delta \theta$ the libration angular amplitude, and $\boldsymbol{e_z}$ is the unit vector in the direction of the rotation axis. {{The resulting dimensionless period of libration is given by $T_{lib}=2\pi/\omega$}}. In the frame rotating at the mean rotation rate, the equations governing the fluid motion are
 \begin{eqnarray}\label{syst:eqn_sl}
  \frac{\partial \boldsymbol{u}}{\partial t}+ (\boldsymbol{u}\cdot\boldsymbol{\nabla})\boldsymbol{u} +2\, \boldsymbol{e_z} \times \boldsymbol{u} & = & -\boldsymbol{\nabla}\Pi+E\boldsymbol{\nabla} ^2\boldsymbol{u},\label{syst:eqn_s1l} \\    
\label{syst:eqn_s2l} \nabla \cdot \boldsymbol{u}  & = &   0,
\end{eqnarray}
\noindent where $E$ is the Ekman number defined by
\begin{equation}
E=\frac{\nu}{\Omega_0\,R^2},
\label{ekman}
\end{equation}
\noindent $\boldsymbol{u}$ is the velocity of the fluid in the rotating frame and $\Pi$ is the reduced pressure taking into account the centrifugal force. The fluid satisfies no-slip boundary conditions on the cylinder walls:
\begin{eqnarray}
   \boldsymbol{u} & = & \epsilon\,\cos(\omega\,t)\,\boldsymbol{e_\theta} \quad \mbox{at}\quad r= 1, \label{condition_cylindre1}\\
    \boldsymbol{u} & = & \epsilon \,r\,\cos(\omega\,t)\,\boldsymbol{e_\theta} \quad \mbox{at}\quad z=\pm\alpha/2, \label{condition_cylindre2}
 \end{eqnarray}
\noindent where $\alpha=H/R$ is the aspect ratio of the cylinder, ($r$, $z$) being the cylindrical radial and vertical coordinates.

In this study we consider the limit of small Ekman number, $EÊ\ll 1$, and we assume that spin-up does not occur in the interior during a libration cycle, i.e. $\omega \gg \sqrt{E}$: this means that in the bulk, at first order, the fluid remains in solid body rotation and does not follow the librating outer boundary.

%%%%  Numerical model 
\subsection{Numerical model}
\label{numericmodel}

Equations (\ref{syst:eqn_s1l}) and (\ref{syst:eqn_s2l}) with the boundary conditions (\ref{condition_cylindre1}) - (\ref{condition_cylindre2}) are solved with the commercial software, Comsol Multiphysics\textsuperscript{\circledR}. Axisymmetric simulations are performed using a standard finite element method in which the numerical grid is composed of two subdomains (see figure \ref{schema_cylinder}): (i) a boundary layer domain 
{{ typically composed of 8 layers of  quadriangular elements all along the outer boundaries and around the axisymmetric axis;
}}
 (ii) a bulk zone with triangular elements. 

 All elements are of standard Lagrange $P1-P2$ type (i.e. linear for the pressure field and quadratic for the velocity field). Note that the finite element method does not induce any particular problem around $r=0$ and that no stabilization technique has been used in this work. {{The temporal solver is the so-called Implicit Differential-Algebraic solver (IDA solver), based on backward differenciating formulas\cite{hindmarsh2005}}}. 

 At each time step the system is solved with the sparse direct linear solver PARDISO\cite{schenk2004}. %SLD 

\begin{center}
\begin{figure}
\begin{center}\includegraphics{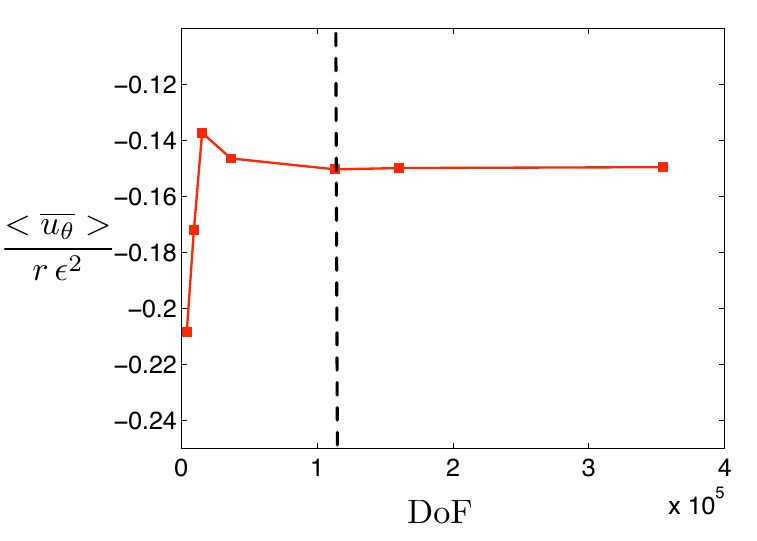}\end{center}
\caption{Convergence with the number of Degrees of Freedom (DoF) of the averaged rotation rate of the zonal flow in the interior, rescaled by $\epsilon^2$ (see section V). The simulations are performed for $E=5.10^{-5}$, $\epsilon=0.1$, $\omega=0.1$ and an aspect ratio $\alpha=2$. The black dashed line corresponds to $113
\,387$ DoF which is used in the following. The value $<\overline{u_\theta}>$ is obtained by averaging over three periods the azimuthal velocity field and then taking the mean value of the zonal flow between $r=0$ and $r=0.5$ at $z=0.2$ which leads to $-0.1496$ for the highest resolution.}
\label{convergence}
\end{figure}
\end{center}

Starting at time $t=0$ with a fluid at rest in the rotating frame, libration of the outer boundary is turned on and computations are pursued during a few spin-up times, $E^{-1/2}$, until a stationary state is obtained.  We then determine the zonal flow $\overline{u_\theta}$ by averaging over three libration periods the numerical velocity field at a fixed axial location. The numerical model has been previously benchmarked and used in Sauret et al.\cite{sauret2010} to study the steady zonal flow induced in a librating sphere in the limit of low libration frequency. The number of degrees of freedom (DoF) used in the simulations is $113\,387$ DoF which is sufficient to ensure accurate results (see the convergence test in figure \ref{convergence}) while keeping a reasonable CPU time {{(i.e. about 8 hours on a standard workstation). Using this method, we systematically explored the range $0.05<\omega<25$ with $0.01<\epsilon<0.85$ and $E$ as small as $E=2\cdot10^{-5}$}}.

%%%%%%%%%%%%%%%%%%%%%%%%%%%%%%%%%%%%%%%%%%%
%%%%%%%%%%%%%%%%  Centrifugal instability %% %%%%%%%%%%%%%
%%%%%%%%%%%%%%%%%%%%%%%%%%%%%%%%%%%%%%%%%%%

\section{Centrifugal instability}

As already documented\cite{noir2010} for fixed values of $\omega$ and $E$, different flow regimes are expected near the sidewall boundary depending on the forcing amplitude $\epsilon$. If $\epsilon$ is smaller than a critical amplitude $\epsilon_c$, the flow remains stable in the whole cylinder (figures \ref{visu}(a) and \ref{visu}(b)). Then, increasing the amplitude $\epsilon$ beyond the critical value $\epsilon_c$ leads to the growth of an instability near the outer boundary (figures \ref{visu}(c) and \ref{visu}(d)). This instability corresponds to a centrifugal destabilization of the time-periodic azimuthal velocity field in the boundary-layer generated by libration. It is characterized by the appearance of counter-rotating Taylor-G\"ortler vortices close to the sidewall boundaries. For larger values of $\epsilon$, the flow becomes more complex (figures \ref{visu}(e) and \ref{visu}(f)). We have not tried to fully characterize this last flow regime because non-axisymmetric features inaccessible to our axisymmetric simulation are then expected to be present (but see section IV).

\begin{center}
\begin{figure}
\begin{center}\includegraphics{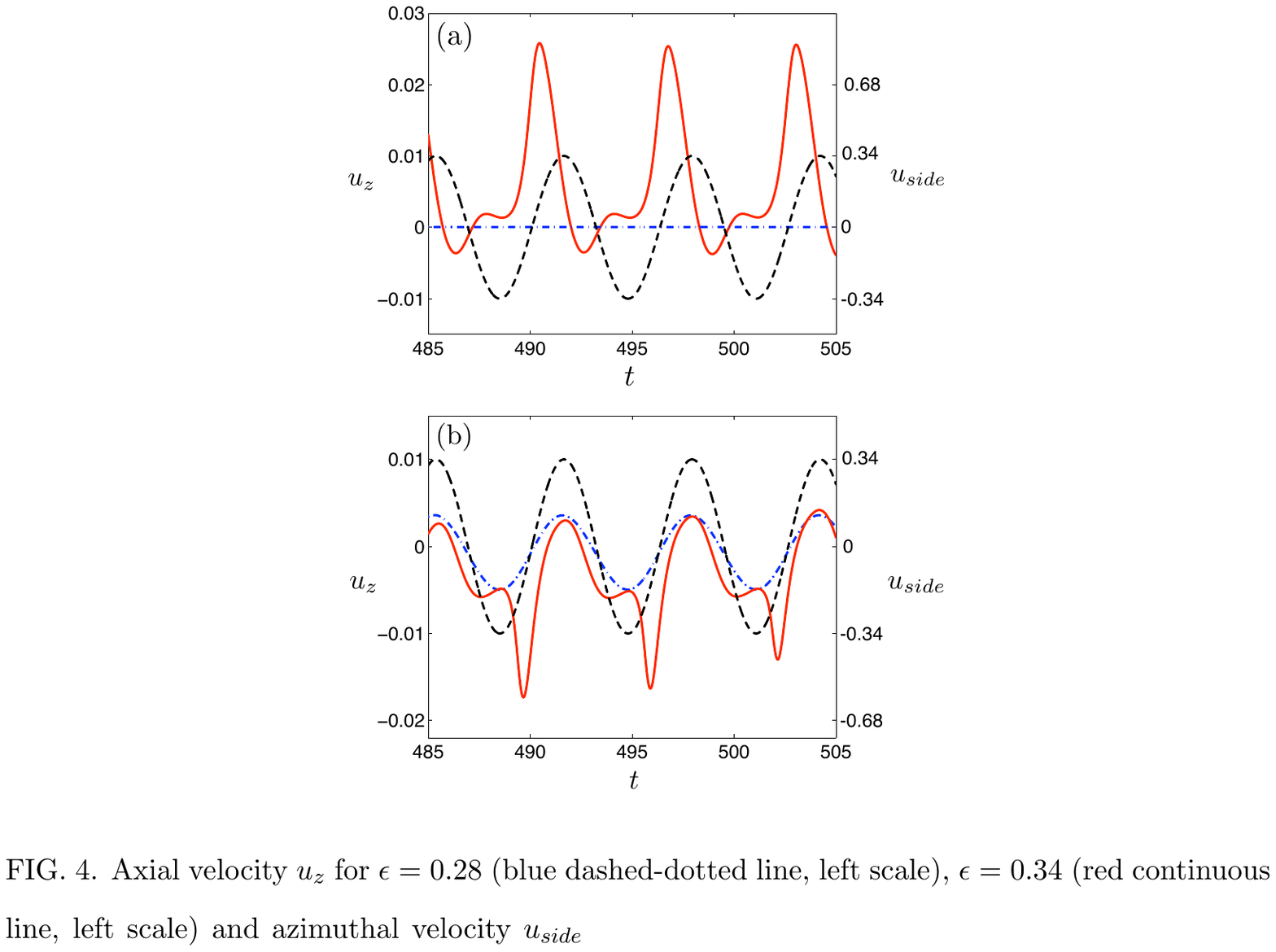}\end{center}
\caption{{{Axial velocity $u_z$ for $\epsilon=0.28$ (blue dashed-dotted line, left scale), $\epsilon=0.34$ (red continuous line, left scale) and azimuthal velocity $u_{side}$ of the sidewall in the rotating frame (black dashed line, right scale) in a cylinder of aspect ratio $\alpha=2$ (a) at the point ($r=0.99$; $z=0$) (b) at the point ($r=0.99$; $z=0.1$). Here, $E=4.10^{-5}$ and $\omega=1$.}}}
\label{profil}
\end{figure}
\end{center}

\begin{widetext}
\begin{center}
\begin{figure}
\begin{center}\includegraphics{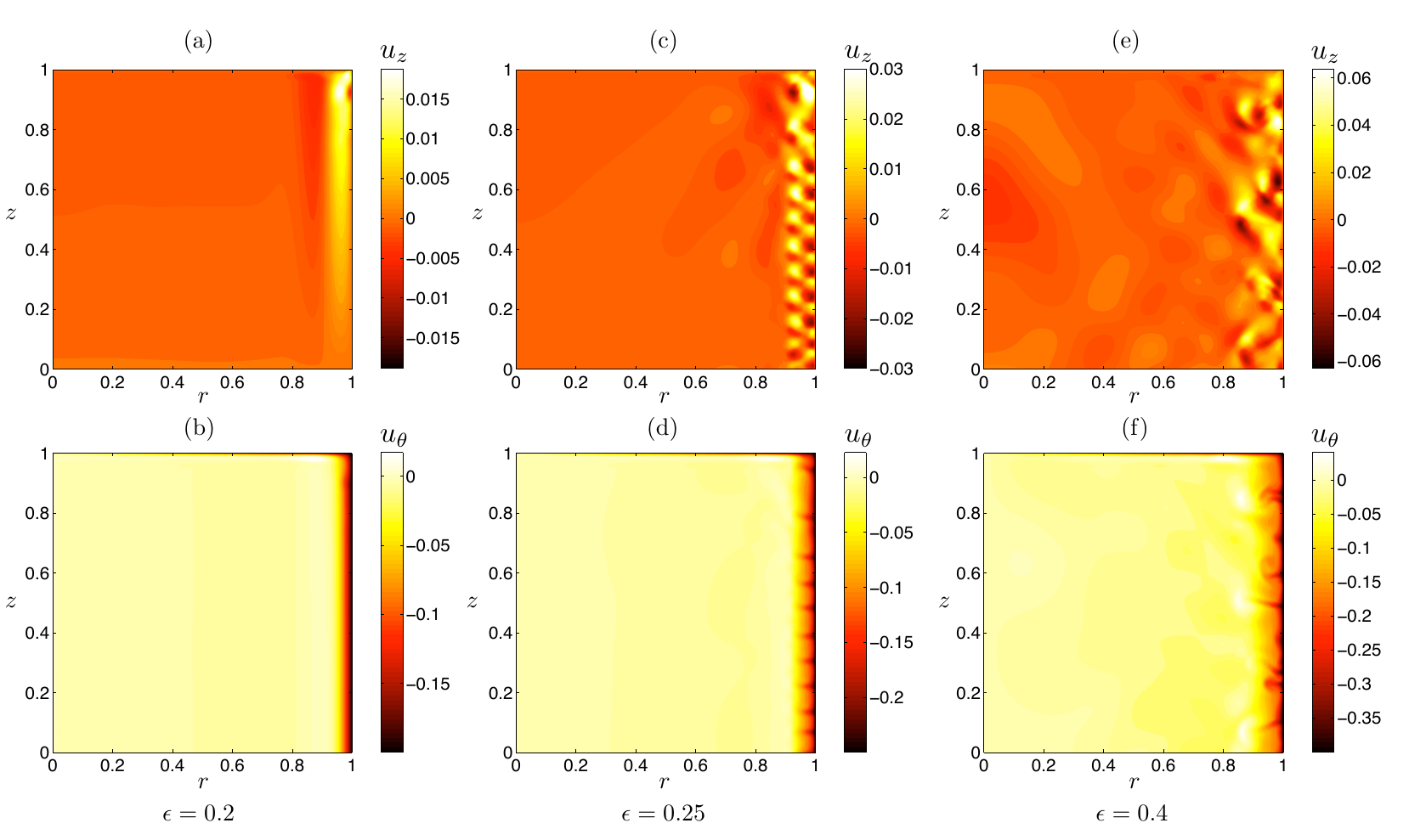}\end{center}
\caption{Visualizations of the libration flow in the upper half of the cylinder ($0<r<1;0<z<1$): axial velocity $u_z$ (upper pictures) and azimuthal velocity $u_\theta$ (lower pictures) at the same time during the retrograde phase ($t=0.4015\,T_{lib}$). All simulations have been done for $\alpha=2$, $E=5.10^{-5}$ and $\omega=0.05$. (a) and (b) show a stable case where no instability is observed. (c) and (d) are centrifugally unstable with the generation of Taylor-G\"ortler vortices. (e) and (f) illustrate the more complex flow that takes place for a larger excitation amplitude $\epsilon$.}
\label{visu}
\end{figure}
\end{center}
\end{widetext}
{{ Figure \ref{profil}(a) and (b) show typical temporal signals of the axial velocity near the sidewall boundary in stable and unstable regimes. In both figures, the azimuthal velocity of the sidewall is also plotted. In figure \ref{profil}(a), the axial velocity is taken at the location $r=0.99$, $z=0$ that is very close to the boundary and on the horizontal plane of symmetry of the cylinder. Below the instability threshold (for $\epsilon=0.28$), no signal is observed at this point whereas a signal oscillation at the libration period is observed at $z=0.1$ (figure \ref{profil}(b)). These oscillations, which are visible when we are not in the plane of symmetry, come from the Ekman pumping from top and bottom boundary layers during the retrograde and prograde phases.}}

{{ When $\epsilon$ is increased above the instability threshold, the axial velocity signal is modified. Either it becomes visible (at $z=0$; figure \ref{profil}(a)) or it reaches larger amplitude (at $z=0.1$; figure \ref{profil}(b)). It can be associated with the growth of the centrifugal instability at the end of the retrograde phase. A net peak is indeed observed in the middle of the prograde phase at both locations.}}

In a previous study, using direct flow visualization with Kalliroscope particles, Noir et al.\cite{noir2009} defined a boundary layer Reynolds number $Re_{BL}$ as the relative strength of inertial and viscous forces in the boundary layer: $Re_{BL}=u\,\delta/\nu$ where $\nu$ is the kinematic viscosity of the fluid, $u$ is the characteristic azimuthal velocity and $\delta$ is the boundary layer thickness. For $Re_{BL}>Re_{BL,cr}$ where $Re_{BL,cr}$ is a critical boundary layer Reynolds number obtained experimentally, a centrifugal instability develops near the sidewall. Using scaling arguments, they established that $Re_{BL,cr}$ should scale as $\epsilon\,E^{-1/2}$ in the spherical case, which was then confirmed numerically by Calkins et al.\cite{calkins2010}. The influence of the libration frequency $\omega$ was not studied. In the cylinder, based on laboratory experiments, Noir et al.\cite{noir2010} suggested that the critical boundary layer Reynolds number should scale as $\epsilon\,E^{-3/4}$. In the following, a theoretical expression for the onset of Taylor-G\"ortler vortices is derived using local stability considerations. This expression takes into account the dependence with the frequency and reconciliates the cylindrical and spherical geometries.

\subsection{Theoretical determination of the threshold of the centrifugal instability}

Let us consider the situation presented in section II. Assuming that the amplitude of excitation is small, $\epsilon \ll 1$, in the rotating frame, one can expand the velocity field in power of $\epsilon$. At the order $\epsilon$, equation (\ref{syst:eqn_sl}) writes
\begin{eqnarray}
\frac{\partial \boldsymbol{u^1}}{\partial t} +2\,\boldsymbol{e_z}\times\boldsymbol{u^1} =  -\boldsymbol{\nabla}p^1+E\,\boldsymbol{\nabla}^2\boldsymbol{u^1}, \label{eq_TG}
\end{eqnarray}

\noindent where the subscript $1$ denotes the terms of first-order in $\epsilon$. By performing a classical boundary layer analysis\cite{wang1970}, we obtain the azimuthal correction of the velocity field in the sidewall boundary layer:
\begin{eqnarray}\label{TG1}
{u_\theta}^1 & = & \epsilon\,\text{exp}\left({-\frac{1-r}{\sqrt{E}}\,\sqrt{\frac{\omega}{2}}}\right)\,\cos\Biggl(\omega t-\frac{1-r}{\sqrt{E}}\,\sqrt{\frac{\omega}{2}}\Biggr). 
\end{eqnarray}

{{ This boundary layer flow of thickness $O(\sqrt{E/\om})$ corresponds to what is called a Stokes layer. 
As long as the period of oscillation is small compared to the spin-up time (that is $\om \gg \sqrt{E}$), this boundary layer flow does not generate any flow in the bulk of the fluid\cite{wang1970,busse2010b}. By contrast, the boundary layer flow on top and bottom boundaries induces a $E^{1/2}$ correction of the flow in both the bulk and the sidewall boundary layer (see Wang\cite{wang1970}).  This weak axial pumping flow remains small when $\om \gg \sqrt{E}$ and is neglected in the present analysis.  
}}

The experimental results \cite{noir2010} tend to demonstrate that the flow (\ref{TG1}) can become centrifugally unstable with respect to axisymmetric perturbations. Is is then natural to apply the Rayleigh criterion, which is a necessary\cite{rayleigh1917} and sufficient\cite{synge1933} condition for instability of axisymmetric inviscid perturbations in a stationary flow. This criterion is
\begin{eqnarray}\label{TG2}
\Phi=\frac{\text{d}}{\text{d}r}\left(r^2\,{u_\theta^{(tot)}}^2\right) <0,
\end{eqnarray}

\noindent where $u_\theta^{(tot)}$ is the total azimuthal velocity field in the absolute frame of reference. This criterion can be satisfied only in the boundary layer where the Rayleigh discriminant 
$\Phi$ reduces at first order to
\begin{eqnarray}\label{TG22222}
\Phi \sim 2\left(2+\frac{\text{d}u_\theta^1}{\text{d}r}\right).
\end{eqnarray}

{{
By using (\ref{TG1}) for $u_{\theta}^1$ in (\ref{TG22222}), we implicitly perform a quasi-steady analysis. In other words, we assume that the time variations
of the flow are slow compared to the temporal scale associated with the centrifugal instability. The instability criterion is thus local in time. By considering 
only the moment where the base flow is the most unstable, we then expect to provide a lower boundary for the instability domain. Here, the most unstable profile is 
reached at $t_m =3\,\pi/(4\,\omega)$ and the minimum value of the Rayleigh discriminant $\Phi$ is obtained at $r_m=1$ and equals:
}}
\begin{eqnarray}\label{temp1}
\Phi_{min}  =  2\left(2-\epsilon\,\sqrt{\frac{\omega}{E}}\right).
\end{eqnarray}

When $\Phi_{min}>0$, the Rayleigh criterion is not satisfied and we thus do not expect instabilities. In terms of $\epsilon$, this means that if  
\begin{eqnarray}
\epsilon < \epsilon_c^{(NV)}=2\,\sqrt{\frac{E}{\omega}},
\end{eqnarray}
\noindent the flow is stable. This criterion is non-viscous and does not take into account the viscous damping that will affect the small scale instability modes. To obtain a viscous criterion, it is necessary to solve the viscous stability equation for the flow given by (\ref{TG1}). Taking into account that the most unstable profile is obtained at $t_m$, a lower limit for the stability threshold can be obtained by analysing the stability properties of such a profile. Assuming both $\epsilon \ll 1$ and $\sqrt{E/\omega}\ll 1$ with $\epsilon\,\sqrt{\omega/E}=O(1)$, the viscous stability problem for axisymmetric perturbations $(u_r,u_\theta,u_z,p)\,\text{exp}\left(\text{i}\,k\,\tilde{z}+\sigma\,t\right)$ reduces to solving the system:
\begin{eqnarray}
\sigma\,u_r-2\,u_\theta=\frac{\partial \tilde{p}}{\partial \tilde{r}}+\frac{\omega}{2}\,\left(\frac{\partial^2}{\partial \tilde{r}^2}-k^2\right)\,u_r, \label{syst_stabilitea}\\
\sigma\,u_\theta+\left(2-\tilde{\epsilon}\,\,\frac{\partial \tilde{u}_\theta^1(\tilde{r})}{\partial \tilde{r}}\right)\,u_r=\frac{\omega}{2}\,\left(\frac{\partial^2}{\partial \tilde{r}^2}-k^2\right)\,u_\theta ,\label{syst_stabiliteb}\\
\sigma\,u_z=-\text{i}\,k\, \tilde{p}+\frac{\omega}{2}\,\left(\frac{\partial^2}{\partial \tilde{r}^2}-k^2\right)\,u_z,\label{syst_stabilitec} \\
- \frac{\partial u_r}{\partial \tilde{r}}+\text{i}\,k\,u_z=0,\label{syst_stabilited}
\end{eqnarray}
\noindent with the boundary conditions
\begin{eqnarray}\label{cond_stabilite}
u_r=u_\theta=u_z= 0 \qquad \text{at} \qquad \tilde{r}=0,+\infty,
\end{eqnarray}

\noindent where
\begin{eqnarray}
\tilde{r}=\frac{1-r}{\sqrt{E}}\,\sqrt{\frac{\omega}{2}}, \\
\tilde{z}=\frac{z}{\sqrt{E}}\,\sqrt{\frac{\omega}{2}}, \\
\tilde{p}=\frac{p}{\sqrt{E}}\,\sqrt{\frac{\omega}{2}}, \\
\tilde{\epsilon}=\frac{\epsilon}{\sqrt{E}}\,\sqrt{\frac{\omega}{2}}, \\
\tilde{u}_\theta^1(\tilde{r})=\text{e}^{-\tilde{r}}\,\sqrt{2}\,\cos(\tilde{r}).
\end{eqnarray}

{{ This eigenvalue problem for $\sigma$  is solved numerically using a pseudospectral method. The radial discretization is based on Chebyshev functions.
}}
The parameter $E$ has disappeared from the equations. This means that the eigenvalues $\sigma$ obtained by solving (\ref{syst_stabilitea})-(\ref{syst_stabilited}) with (\ref{cond_stabilite}) are function of $k$, $\tilde{\epsilon}$ and $\omega$ only. This implies that the maximum growth rate over all possible $k$ is a function of $\tilde{\epsilon}$ and $\omega$ only:
\begin{equation}
 \sigma_{max}=\displaystyle \max_{k} \sigma=\sigma_{max}(\tilde{\epsilon},\omega)
\end{equation}

The stability curve, defined by $\sigma_{max}=0$, is then given by a curve $\tilde{\epsilon}=\tilde{\epsilon}_c(\omega)$ which is a function of $\omega$ only. Using the primitive variable, the viscous criterion for stability therefore reads
\begin{equation}\label{viscouscriterion}
\epsilon < \epsilon_c^{(V)}=\tilde{\epsilon}_c(\omega)\,\sqrt{\frac{2\,E}{\omega}}.
\end{equation} 

\begin{center}
\begin{figure}
\begin{center}\includegraphics{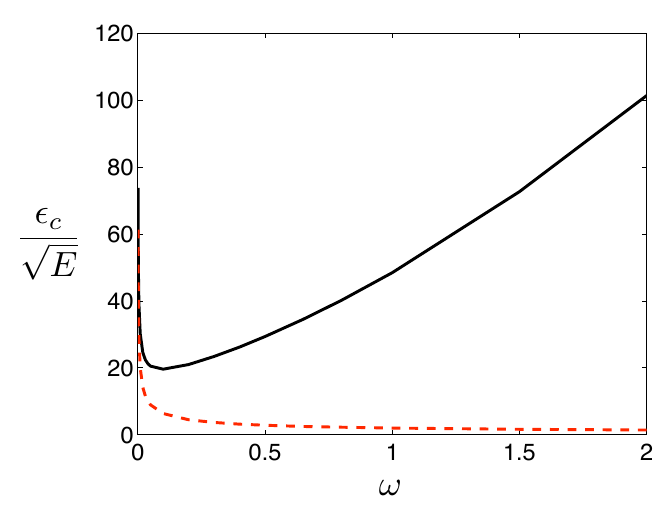}\end{center}
\caption{Critical amplitude $\epsilon_c/\sqrt{E}$  for the centrifugal instability obtained by the theory. Solid line: viscous criterion $\epsilon_c^{(V)}(\omega)/\sqrt{E}$; dashed line: inviscid criterion $\epsilon_c^{(NV)}(\omega)/\sqrt{E}={2}/{\sqrt{\omega}}$}
\label{epsilonc}
\end{figure}
\end{center}

In figure \ref{epsilonc}, we have plotted $\epsilon_c^{(V)}(\omega)/\sqrt{E}$ as a function of $\omega$. This plot is compared to the non viscous prediction $\epsilon_c^{(NV)}(\omega)/\sqrt{E}={2}/{\sqrt{\omega}}$. Note that when $\omega \to 0$, $\epsilon_c^{(V)} \sim \epsilon_c^{(NV)}$. This is in agreement with the system of equation (\ref{syst_stabilitea})-(\ref{syst_stabilited}) in which the viscous terms vanish as $\omega \to 0$. However, both $\epsilon_c^{(NV)}$ and $\epsilon_c^{(V)}$ diverge as $\omega \to 0$. This divergence is associated with the blow up of boundary layer thickness when $\omega \to 0$. In fact, this boundary layer does not blow up but transforms into a Stewartson layer of thickness $E^{1/4}$ when $\omega$ becomes of order $E^{1/2}$ or smaller. Under the hypothesis $\omega \gg \sqrt{E}$ we never reach this regime. Moreover, we see that neglecting the viscosity leads to a non-viscous criterion which decreases while increasing $\omega$, in contradiction with the experiments (see below). On the contrary, the viscous criterion has a correct tendency to increase with $\omega$.

The above viscous stability criterion has been obtained by considering the most unstable profile in a frozen state. It should then, a priori, provide a lower limit for the critical destabilization amplitude. In principle, as the base flow oscillates, a Floquet stability analysis should be performed to obtain a more precise estimate for the critical amplitude but we discuss below why we do not think that such a development is worthwhile. Finally, note that the above viscous stability criterion has only assumed that the boundary layer is locally vertical. It should therefore apply close to the vertical boundary of any axisymmetric containers. In particular, we expect the same criterion to apply for the flow in a sphere near the equator. In the following sections, the viscous stability criterion is compared to the numerical simulations and experimental observations.

\subsection{Influence of the Ekman number}

\begin{center}
\begin{figure}
\begin{center}\includegraphics{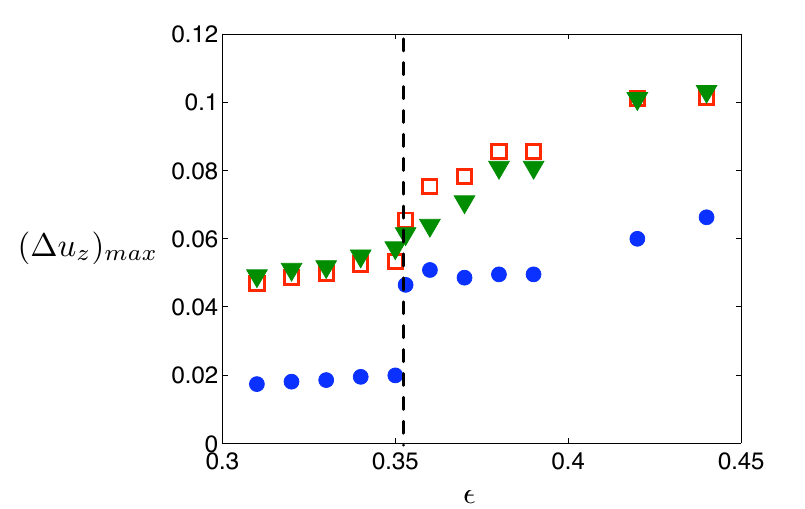}\end{center}
\caption{Amplitude of the axial peak-to-peak velocity $(\Delta u_z)_{max}$ in a cylinder of aspect ratio $\alpha=2$ for $E=6.10^{-5}$, $\omega=0.8$ and various values of $\epsilon$. The measurements are performed at different locations: the blue circles at $(r=0.95;z=0.1)$, the red squares at $(r=0.95;z=0.4)$ and the green triangles at $(r=0.95;z=0.9)$. We see a clear transition between stable/unstable regimes at $\epsilon_c =0.355$.}
\label{TGstab}
\end{figure}
\end{center}

Our numerical simulations allow us to recover the range of parameters explored in previous experiments in the literature\cite{noir2010} but also to increase the range of studied Ekman numbers. Moreover we can explore more precisely the transition between the stable/unstable regimes with the measurement of the axial velocity. Finally, we can study the influence of the libration frequency which was not investigated previously. The method used to determine the stability of the flow is the same as the one used by Calkins et al.\cite{calkins2010}: for given $E$ and $\omega$ we start at an arbitrary amplitude of libration $\epsilon$ and observe if the flow is stable (respectively unstable), then we increase (respectively decrease) the amplitude of libration until we reach an unstable (respectively stable) regime. The determination of the instability is done both by direct visualization, and by plotting the vertical velocity near the outer boundary at some locations as it has been observed that an important variation of its amplitude is visible at the transition between stable and centrifugally unstable regimes. If the flow is stable, the vertical velocity remains small and its amplitude is periodic with a period equal to the libration period (see figure \ref{profil}(a)). Once we reach the unstable regime the vertical velocity suddenly increases (figure \ref{TGstab}) and becomes less regular (see figure \ref{profil}(b)). The critical libration amplitude $\epsilon_c$ is defined as the largest stable $\epsilon$. We have not observed any hysteresis. We therefore expect the instability to be supercritical as for classical Taylor-Couette flow.

Figure \ref{TGstab} shows the maximum peak to peak amplitude, i.e. the maximum value minus the minimum value, of the vertical velocity as a function of the amplitude of excitation $\epsilon$ for a given Ekman number and a given libration frequency. Due to the presence of Ekman pumping from top and bottom boundaries, even below threshold, the value of the axial velocity depends on the location of the measurement. The axial velocity is weak close to the plane of symmetry but large close to the top and bottom boundaries. A transition is visible close to $\epsilon_c=0.355$. This transition is very clear in the measurements at $z=0.1$ as the axial velocity increases by a factor three. Visualizations also show that the unstable rolls do not always develop at the same location in the cylinder. For small libration frequencies, typically $\omega <1$, we observe that the Taylor-G\"ortler vortices first develop near the equator of the cylinder, i.e. at $z \simeq 0$. Increasing the frequency of libration leads to the first development of these unstable rolls at mid-distance between the corner and the middle plane. This feature is probably also related to the Ekman pumping from top and bottom boundaries.

\begin{center}
\begin{figure}
\begin{center}\includegraphics{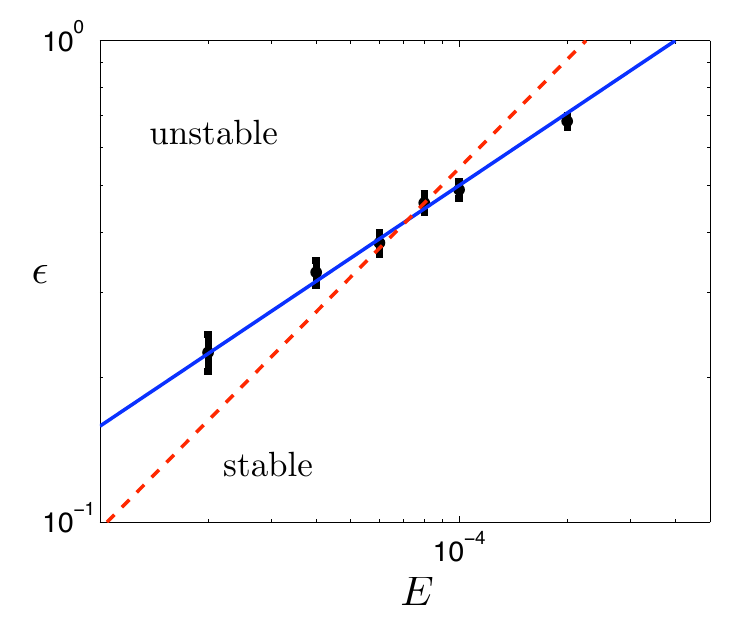}\end{center}
\caption{Stability diagram for the onset of Taylor-G\"ortler vortices for a libration frequency of $\omega=1$ in a cylinder of aspect ratio $\alpha=2$. The black bars are the numerical results together with the uncertainty in the estimation of the threshold. The blue continuous line is the viscous stability criterion (\ref{viscouscriterion}) without adjusting parameter: $47.8\,E^{1/2}$ and the red dashed line is the best fit line using the scaling of Noir et al.\cite{noir2010}: $543.2\,E^{3/4}$.}
\label{numericsyst}
\end{figure}
\end{center}

Figure \ref{numericsyst} presents the results of our numerical investigation of the Ekman number dependence. The $E^{1/2}$-scaling law is clearly observed and is well fitted by the viscous stability threshold (\ref{viscouscriterion}) which gives $\epsilon_c = 47.8\,E^{1/2}$ at $\omega=1$ with no adjusting parameter. {{If we take $\delta= \sqrt{E}\sqrt{2\Omega_0/\omega} R$ as spatial scale and $u=\sqrt{2}\epsilon \Omega_0 R/2$ as  velocity scale, as prescribed by the boundary layer theory (see (\ref{TG1})), }}
this corresponds to a critical boundary layer Reynolds number $Re_{BL,cr}=47.8$. This is surprisingly close to the value $Re_{BL,cr}=46$ obtained in a spherical shell by Calkins et al.\cite{calkins2010} at the same $\omega$.

\begin{center}
\begin{figure}
\begin{center}\includegraphics{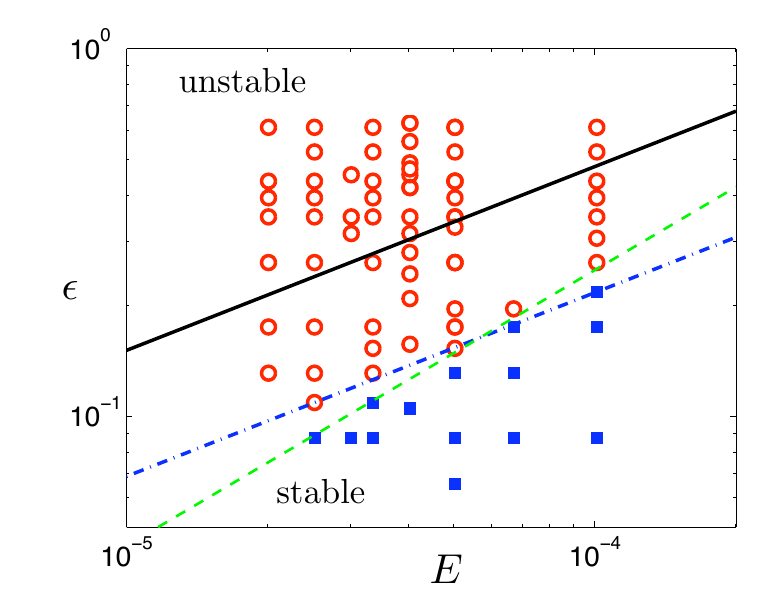}\end{center}
\caption{Experimental results from Noir et al.\cite{noir2010} obtained by direct visualization plotted in terms of the libration forcing $\epsilon$ versus $E$ for a fixed value of $\omega=1$. The blue squares are centrifugally stable and the red circles are centrifugally unstable. The black line is the viscous stability criterion (\ref{viscouscriterion}), the blue dashed-dotted line is the best fit line in $E^{1/2}$: $22\,E^{1/2}$, and the green dashed line is the fit line suggested by Noir et al.\cite{noir2010}: $250\,E^{3/4}$.}
\label{noir}
\end{figure}
\end{center}

In figure \ref{noir}, we have reported the experimental observations of Noir et al.\cite{noir2009} for the same aspect ratio as in figure \ref{numericsyst}. The theoretical prediction {{and the scaling law suggested by Noir 
et al.\cite{noir2010} are also shown}}. The agreement with the theory is not as good as with our numerical simulations. In particular we do not understand the fact that the experimental threshold is well below our estimate which is expected to be a lower limit. The same observation was made in the sphere between experimental\cite{noir2009} and numerical\cite{calkins2010} results. The discrepancy between these measurements was explained by Calkins et al.\cite{calkins2010} by a systematic higher noise level present in the experimental setup. It could also be due to the visualization with Kalliroscope particles which do not directly measure the velocity field. Note that a scaling in $E^{1/2}$ still fits the experimental data with a good agreement. {{Therefore, regarding the experimental measurement both scalings in $E^{1/2}$ or $E^{3/4}$ could a priori be possible}}. However, the  latter scaling in $E^{3/4}$ does not hold for the numerical results (see figure \ref{numericsyst}).

\subsection{Influence of the libration frequency $\omega$}

\begin{center}
\begin{figure}
\begin{center}\includegraphics{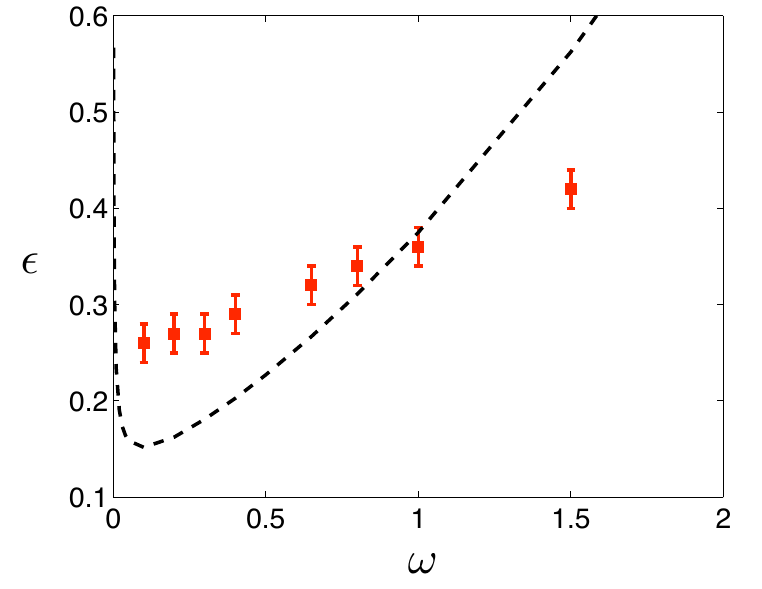}\end{center}
\caption{Stability diagram for the onset of Taylor-G\"ortler vortices as a function of the frequency of libration $\omega$. The squares are the numerical simulations for $E=6.10^{-5}$ in a cylinder of aspect ratio $\alpha=2$ and the dashed line is the viscous stability criterion (\ref{viscouscriterion}) with no adjustment parameter.}
\label{TGomega}
\end{figure}
\end{center}

The influence of the libration frequency $\omega$ on the instability threshold has not been considered in previous studies. Here we consider $8$ different values of {{$\omega$ arbitrarily chosen in the range $\omega \in[0;2]$: $0.1$, $0.2$, $0.3$, $0.4$, $0.65$, $0.8$, $1$ and $1.5$}}. The results are reported in figure \ref{TGomega}. A clear increase of the instability threshold with $\omega$ is observed in the simulation. In this figure, we have also plotted the viscous stability criterion (\ref{viscouscriterion}). Whereas the theoretical prediction captures correctly the global increase of the threshold with $\omega$, it fails in providing correct quantitative values for small and large $\omega$.

For small $\omega$, we predict a threshold which is twice smaller than the numerically observed value. We cannot attribute this difference to the time variation of the base flow, which was not taken into account  in the theory. Indeed for small $\omega$, the quasi-stationnary approximation applies and we expect a Floquet stability analysis to provide a similar stability threshold. Instead, we suspect the difference to be associated with a bad description of the boundary layer flow.
{{ 
For small frequencies (if the condition $\omega \gg \sqrt{E}$ is no longer satisfied), transient  Stewartson layers generated by top and bottom boundaries could have time to appear \cite{barcilon1968} and modify the Stokes layer solution
(\ref{TG1}). }}

For large $\omega$, our theoretical prediction is above the numerical stability threshold. Again, a more precise Floquet analysis would have therefore not explained the difference.
{{ Indeed, the global stability results obtained by a Floquet analysis would have provided a smaller instability domain and therefore an even larger value for the threshold. }}
Here we think that the discrepencies come from the nonlinear effects. Our theoretical approach which assumes a linear boundary layer flow is indeed expected to break down when $\epsilon$ is not sufficiently small.

%%%%%%%%%%%%%%%%%%%%%%%%%%%%%%%%%%%%%%%%%%%
%%%%%%%  Spontaneous generation of inertial modes %%%%%%%%%%%
%%%%%%%%%%%%%%%%%%%%%%%%%%%%%%%%%%%%%%%%%%%

\section{Spontaneous generation of inertial modes}

In the previous section, we have studied the occurrence of the centrifugal instability on the sidewalls and have mainly considered the instability threshold. In this section, we consider configurations well above the threshold for which rich dynamics are observed. Our numerical simulations tend to demonstrate that increasing the libration amplitude $\epsilon$ eventually leads to a turbulent flow in the boundary layer. We have also observed that these turbulent regimes can drive motion in the bulk under the form of spontaneously generated inertial waves.

\begin{figure}
\begin{center}
\includegraphics{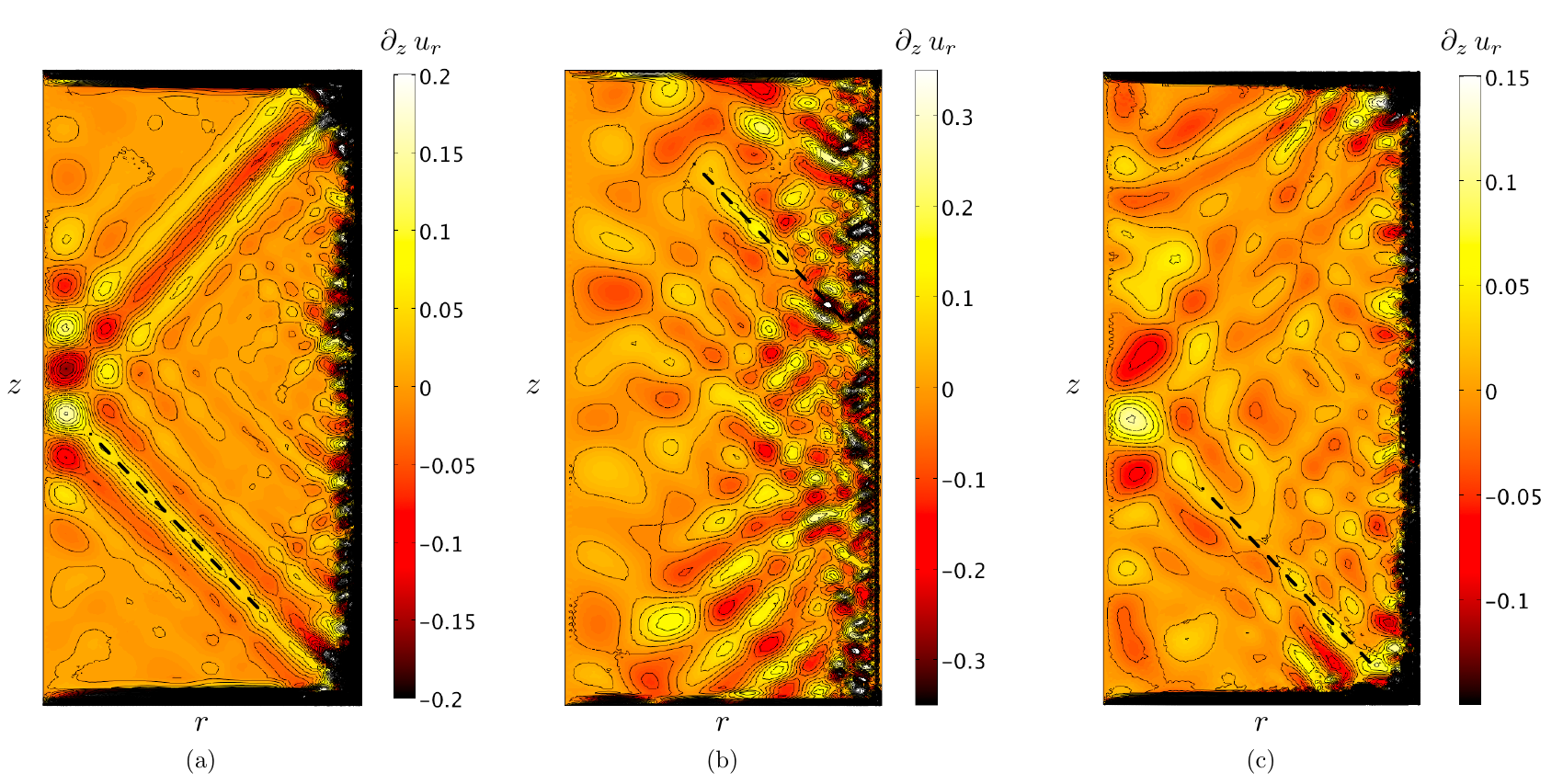}
\caption{{Axial derivative of the radial velocity in the librating cylinder of aspect ratio $\alpha=2$ at time $t=3\pi/(2\,\omega)$ for (a) $\omega=2.5$, $\epsilon=0.70$, (b) $\omega=0.1$, $\epsilon=0.6$, (c) $\omega=4.1$, $\epsilon=0.85$. All simulations are performed for $E=3.10^{-5}$. The amplitude of the colorbar is chosen to visualize inertial modes (a saturation of the axial velocity near the outer boundary of the container is thus visible). Dotted lines correspond to an orientation of $45\char23$}.}
\label{generation_inertial2}
\end{center}
\end{figure}

\begin{figure}
  \begin{center}
      \includegraphics{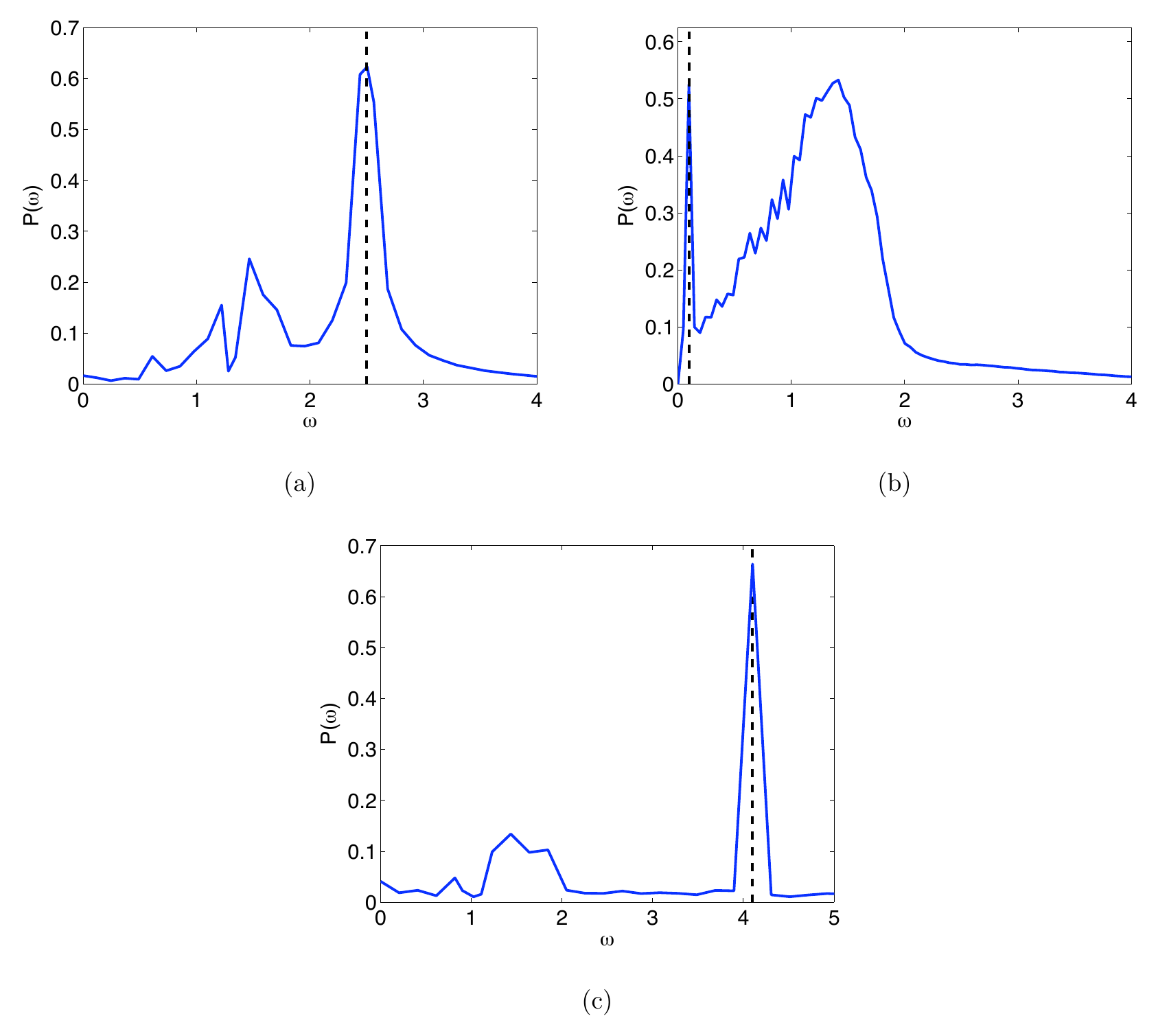} 
\caption{{{Frequency spectrum in a cylinder of aspect ratio $\alpha=2$ for $E=3.10^{-5}$. The three cases corresponds to the situation presented in figure  \ref{generation_inertial2}, namely for (a) $\omega=2.5$, (b) $\omega=0.1$, (c) $\omega=4.1$. The spectrum is obtained by axially averaging the radial velocity at $r=0.6$. The black dotted line indicates in each cases the libration frequency}}}
    \label{fig:spectrum}
  \end{center}
\end{figure} 

{{In figure \ref{generation_inertial2}, we have plotted $\partial_z\,u_r$ for configurations above the instability threshold for different frequencies. In all cases, we are well above the threshold, i.e. in regimes where the boundary layer can be considered as turbulent. We observe structures in the bulk of the fluid,
  which seem to share some common features.  
In particular, we observe in all cases, small scale structures oriented at a fixed angle  $\theta \approx 45\char23$. 
We think that these structures are inertial waves excited by the turbulent motion  in the boundary layer.  
In figure \ref{fig:spectrum}, we have plotted a typical frequency spectrum obtained for each case.  
For each configuration, we do observe a peak at the frequency of excitation, that is $\om=2.5$ in (a), $\om=0.1$ in  (b),
and $\om=4.1$ in (c).
Each case also possesses a peak at approximatively $\om \approx 1.4$ which corresponds to the frequency of the inertial wave pattern oriented at $45\!\char23\!\!$. 
Indeed, the  orientation pattern is also the direction of the group velocity of the inertial wave packet which is related to its frequency 
by the relation\cite{greenspanbook} $\om = 2 \sin \theta $. This relation gives $\om \approx 1.41$ for  $\theta \approx 45\!\char23\!\!$.
}} 

{{The case (a) of a forcing frequency $\omega=2.5$ was previously studied by Lopez \& Marques \cite{lopez2011}. They  also found coherent structures in the bulk. They  argued that those structures are inertial modes oscillating at $\omega/2$ which are excited by a period-doubling mechanism. 
Here this mechanism gives a frequency $\omega_{ex}=1.25$ which corresponds to the second peak in figure \ref{fig:spectrum}(a). The pattern observed in figure \ref{generation_inertial2}(a),
especially the two inertial wave beams  emitted from the corners, does not correspond to this inertial mode. We think that it is associated with a stronger turbulence activity near the corners
for this particular forcing frequency.  }}

{{The case (b) of a forcing frequency $\omega=0.1$ corresponds to the situation where the inertial wave pattern is
the most visible and the most regular. The frequency spectrum  also possesses a larger peak at $\omega \approx 1.4$ in that case.  
Note that other peaks (at $\om=0.2, 0.3, ..$) corresponding to harmonic frequencies of the forcing  are also visible on the spectrum. }}

{{The case (c) of a forcing frequency $\omega=4.1$ is interesting as the period doubling mechanism proposed by Lopez \& Marques can not be active since $\om/2$ is not in the inertial frequency range $(0,2)$.  The inertial wave pattern can not therefore be directly related to the forcing frequency. As for the two other frequencies, a 
pattern preferentially oriented at $45\!\char23$  is still  observed. }}

{{The three different cases have been simulated for different libration frequencies, $\omega=0.1$, $\omega=2.5$ and $\omega=4.1$. Independently of the frequency of the forcing, the inertial waves generated by the turbulent flow near the outer boundary always exhibit a peak in the frequency spectrum around $\omega \simeq 1.4$. It suggests that another mechanism than direct forcing or period doubling could exist for the generation of inertial waves in a cylinder. 
Because the characteristic of the inertial waves do not depend on the forcing frequency, we suspect that the generation could be due to the turbulent motion in the
boundary layer. Such a forcing by turbulence has not been studied in detail in the context of rotating flows. Nevertheless, we can reasonably think that the
results obtained in the context of gravity waves where similar observations were made could apply (see Dohan \& Sutherland\cite{dohan2005}). More works on this interesting issue which is beyond the scope of the present paper are clearly needed.}}

%%%%%%%%%%%%%%%%%%%%%%%%%%%%%%%%%%%%%%%%%%%
%%%%%%%%%%%%%%%%%%%  Zonal flow%% %%%%%%%%%%%%%%%%
%%%%%%%%%%%%%%%%%%%%%%%%%%%%%%%%%%%%%%%%%%%

\section{Mean zonal flow induced by the librational forcing}

In this section we analyse the mean zonal flow induced by the librational forcing that is the time averaged value of the azimuthal velocity. We first consider configurations where inertial modes are not excited in order to compare our numerical results to a previous analytical solution given by Wang\cite{wang1970}. Then we analyse the mean zonal flow when inertial modes are present in the interior and describe the key features of the resulting flow in this regime. 
{{ We assume that the amplitude of libration is sufficiently small such that the sidewall boundary layer remains centrifugally stable. 
}}

%%%%%%%%%%%%%%%%%%%%%%%%%%%%%%%%%%%%%%%%%%
\subsection{Zonal flow for $\boldsymbol{\omega \ll1}$ or $\boldsymbol{\omega>2}$}

A full analytical solution of the zonal flow in a librating cylinder has been obtained by Wang\cite{wang1970} for an arbitrary libration frequency $\omega$, assuming that no inertial mode is excited in the interior. He showed that the mean flow correction is at leading order an azimuthal flow of the form $\overline{u_\theta}=\epsilon^2\,r\,\Omega_2(\omega)$, where $\Omega_2$ is uniform in the bulk of the fluid. The explicit expression of $\Omega_2$ in terms of the frequency is given in the appendix.

Wang \cite{wang1970}  predicted that this mean flow should match the evolution of the sidewall boundary where $\overline{u_\theta}=0$ in a $E^{1/4}$ viscous layer. 

{{  Note that  Wang\cite{wang1970} also demonstrated that a $E^{1/3}$ viscous layer is needed to satisfy the continuity of the axial velocity field. A composite approximation for the azimuthal velocity, which takes into
account the $E^{1/4}$ layer solution, is obtained in the form }}%SLD
\begin{eqnarray}\label{wangcomposite}
\overline{u_\theta} & =& \epsilon^2\,r\,\Omega_2\,\left[1-\text{exp}\left(-\frac{1-r}{E^{1/4}}\right)\right].
\end{eqnarray}
It is this expression, with $\Omega_2$ given by (\ref{wangresult1}) and (\ref{wangresult2}) from the appendix, that we shall test below.

\begin{center}
\begin{figure}
\begin{center}\includegraphics{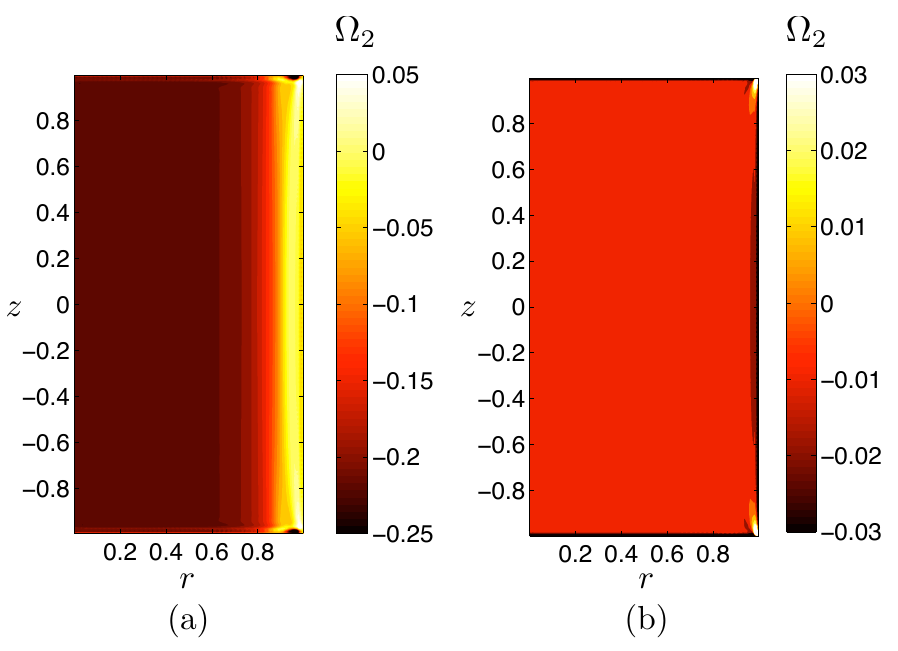}\end{center}
\caption{Angular velocity $\Omega_2=\overline{u_\theta}/(r\,\epsilon^2)$ of the zonal flow in the whole cylinder of aspect ratio $\alpha=2$ obtained numerically for $\epsilon=0.05$, $E=4.10^{-5}$, (a) $\omega=0.05$ and (b) $\omega=6$. The zonal flow is obtained by averaging the azimuthal velocity over 3 libration periods.}\label{2D_noinertial}
\end{figure}
\end{center}

In figure \ref{2D_noinertial}, the mean angular velocity obtained from the numerical simulation is plotted in a meridional plane ($r,z$) for two libration frequencies. We clearly see in these plots that the rotation is uniform in the bulk in both cases. Moreover, near the sidewall ($r=1$), the mean angular velocity is primarilly axially independent on $z$, in qualitative agreement with formula (\ref{wangcomposite}). Note however that the small localized structures in the corners are not described by Wang's theory.

\begin{center}
\begin{figure}
\begin{center}\includegraphics{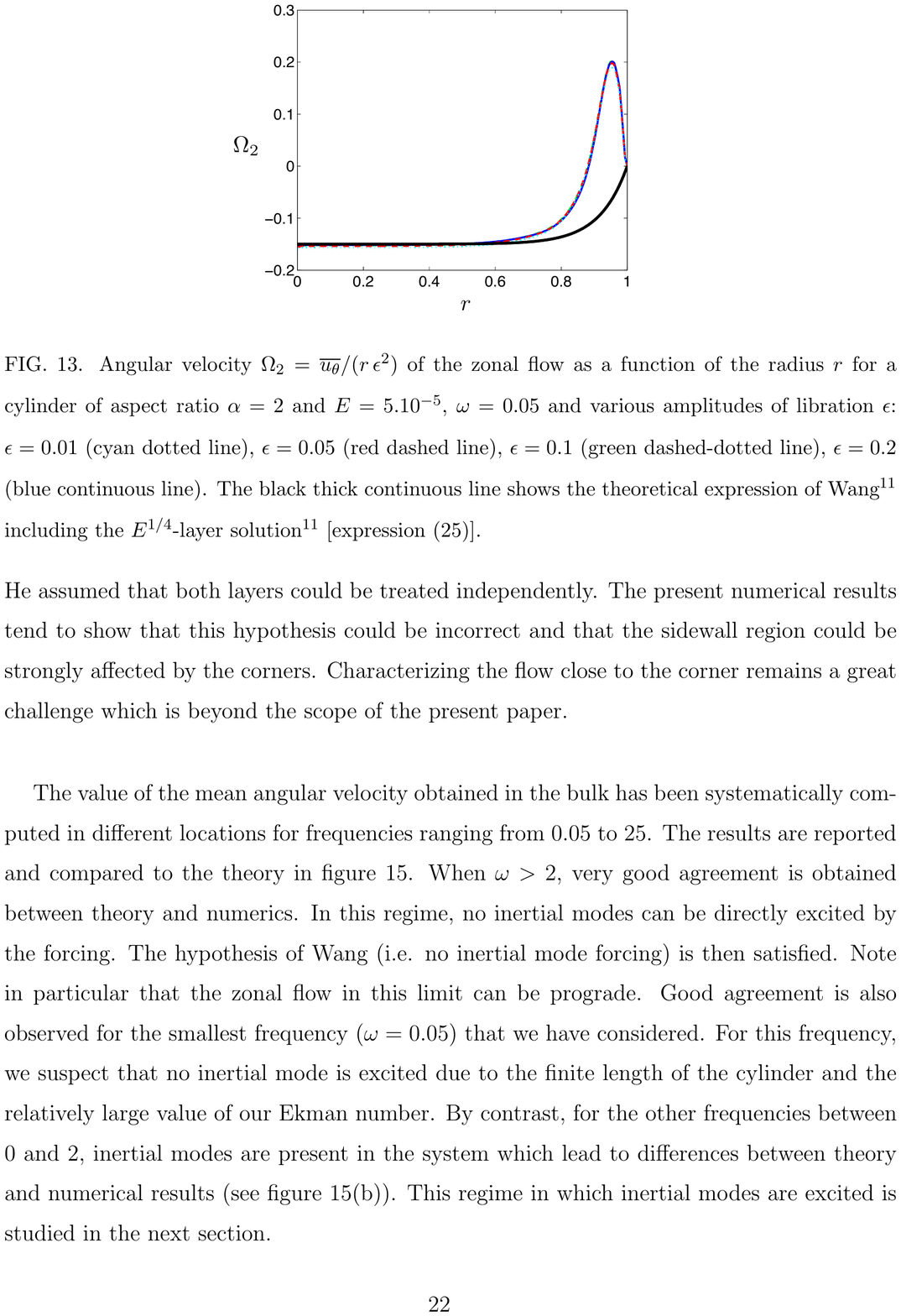}\end{center}
\caption{Angular velocity $\Omega_2=\overline{u_\theta}/(r\,\epsilon^2)$ of the zonal flow as a function of the radius $r$ for a cylinder of aspect ratio $\alpha=2$ and $E=5.10^{-5}$, $\omega=0.05$ and various amplitudes of libration $\epsilon$: $\epsilon=0.01$ (cyan dotted line), $\epsilon=0.05$ (red dashed line), $\epsilon=0.1$ (green dashed-dotted line), $\epsilon=0.2$ (blue continuous line). The black thick continuous line shows the theoretical expression of Wang\cite{wang1970} including the $E^{1/4}$-layer solution\cite{wang1970} [expression (\ref{wangcomposite})].}\label{syst_omegaa}
\end{figure}
\end{center}

\begin{center}
\begin{figure}
\begin{center}\includegraphics{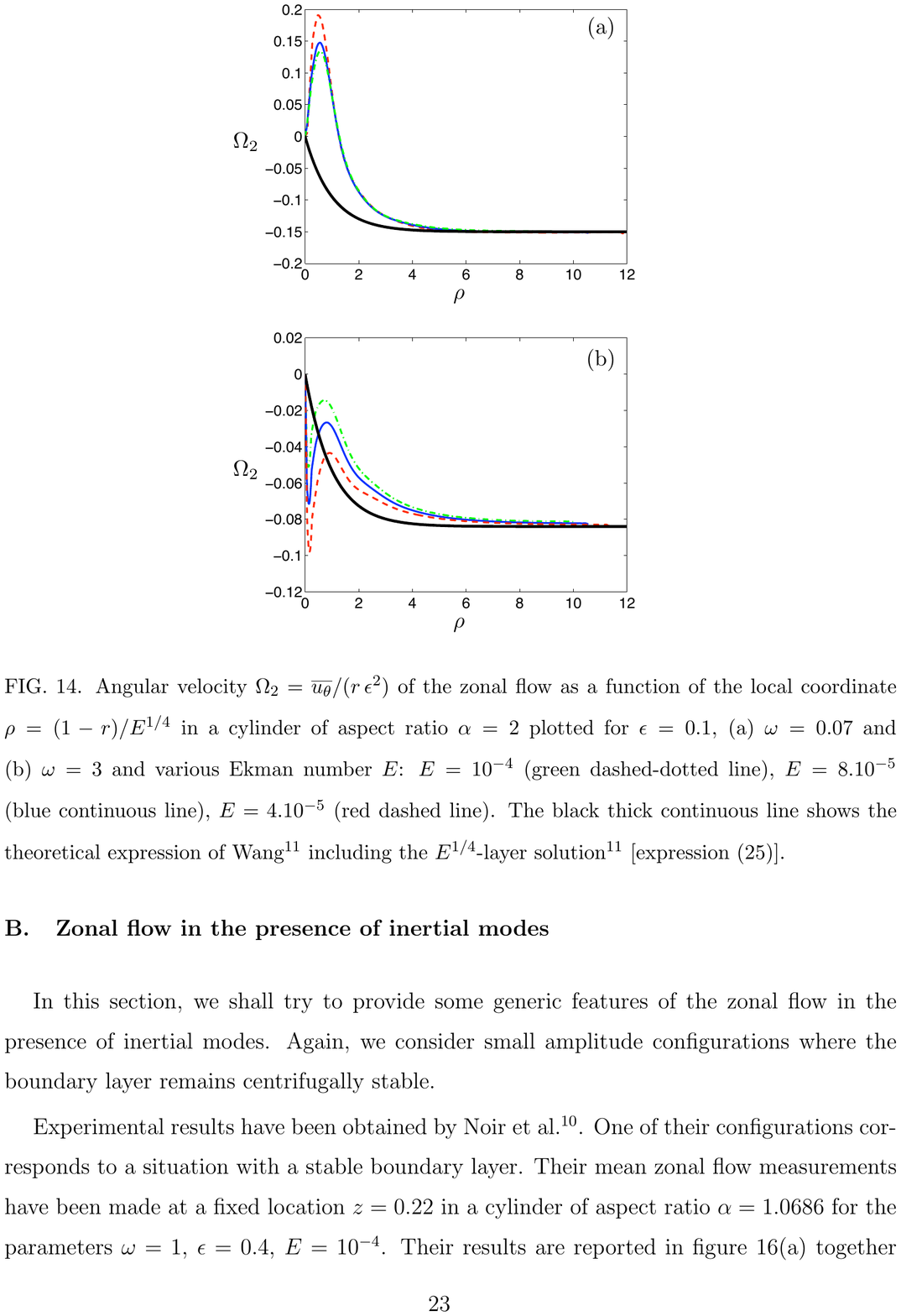}\end{center}
\caption{Angular velocity $\Omega_2=\overline{u_\theta}/(r\,\epsilon^2)$ of the zonal flow as a function of the local coordinate $\rho=(1-r)/E^{1/4}$ in a cylinder of aspect ratio $\alpha=2$ plotted for $\epsilon=0.1$, (a) $\omega=0.07$ and (b) $\omega=3$ and various Ekman number $E$: $E=10^{-4}$ (green dashed-dotted line), $E=8.10^{-5}$ (blue continuous line), $E=4.10^{-5}$ (red dashed line). The black thick continuous line shows the theoretical expression of Wang\cite{wang1970} including the $E^{1/4}$-layer solution\cite{wang1970} [expression (\ref{wangcomposite})].}\label{syst_ekman}
\end{figure}
\end{center}

Figure \ref{syst_omegaa} shows the dependence of the mean angular velocity with respect to $\epsilon$: all the numerical curves for $\Omega_2=\overline{u_\theta}/(r\,\epsilon^2)$  superimpose as $\epsilon$ is varied which demonstrates the scaling in $\epsilon^2$ of the mean zonal flow. In figure \ref{syst_ekman}, we have plotted the rescaled angular velocity $\Omega_2=\overline{u_\theta}/(r\,\epsilon^2)$  as a function of $\rho=(1-r)/E^{1/4}$ keeping $\epsilon$ and $\omega$ fixed but varying the Ekman number for two libration frequencies. We see that for large $\rho$ (that is far from the sidewall) all the curves collapse approximatively to the theoretical prediction. This means that the mean angular velocity in the bulk does not depend on the Ekman number as predicted by the theory. Close to the wall, discrepancies between the theory and the numerics are by contrast visible: in the absence of instabilities, a prograde peak whose amplitude varies with the Ekman number is obtained. 
{{ 
We suspect that the small scale structures observed near the corners are at the origin of these discrepancies. Indeed, Wang\cite{wang1970} did not solve the corner region where sidewall and top/bottom boundary layers 
encounter. He assumed that both layers could be treated independently. The present numerical results tend to show that this hypothesis could be incorrect and that the sidewall region could be strongly affected
by the corners. Characterizing the flow close to the corner remains a great challenge which is beyond the scope of the present paper. 
}}

\begin{center}
\begin{figure}
\begin{center}\includegraphics{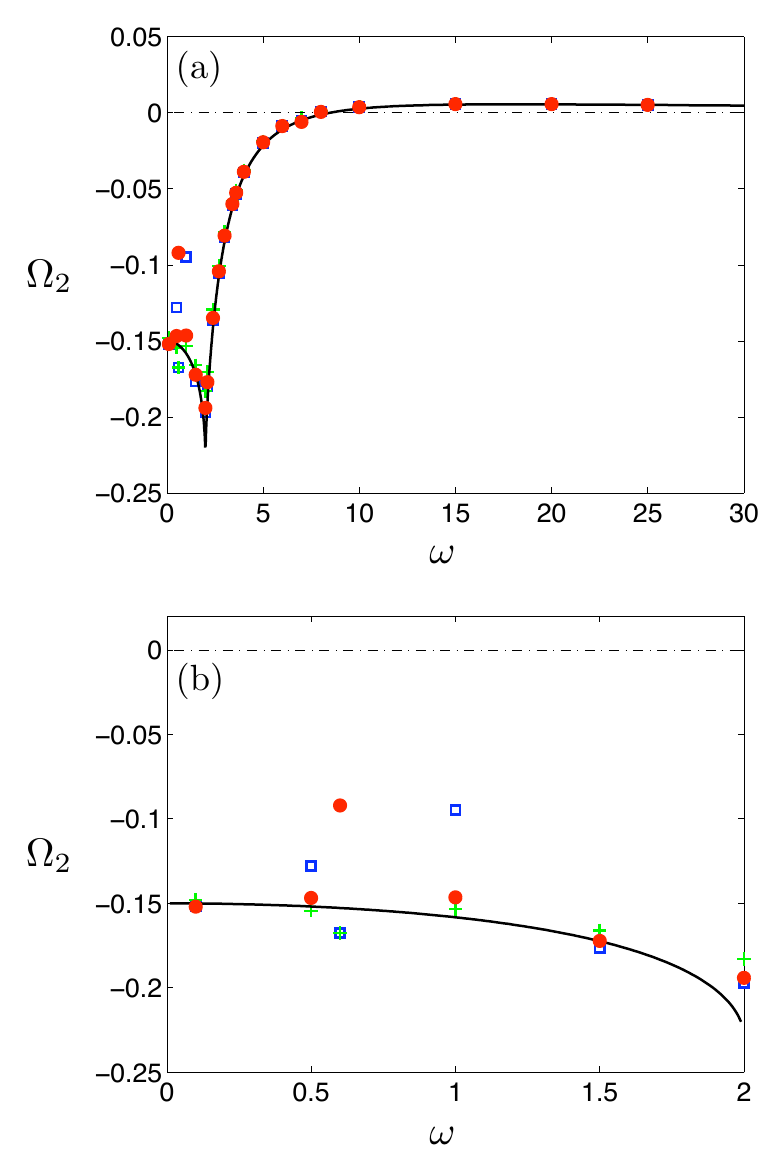}\end{center}
\caption{(a) Angular velocity $\Omega_2=\overline{u_\theta}/(r\,\epsilon^2)$ of the zonal flow, as a function of the frequency of libration in a cylinder of aspect ratio $\alpha=2$ for $E=4.10^{-5}$ and $\epsilon=0.05$. The black continuous line is Wang's prediction [expression (\ref{wangresult1}) or (\ref{wangresult2})], the blue squares are the numerical results at $(r=0.2;z=0.2)$, the red circles at $(r=0.4;z=0.2)$ and the green stars at $(r=0.6;z=0.2)$. The discrepancy for $\omega \in[0.1;2]$ is likely due to the contribution of inertial modes. (b) is a zoom of (a) within the region were inertial modes are expected.} \label{systematique_cylindre}
\end{figure}
\end{center}

The value of the mean angular velocity obtained in the bulk has been systematically computed in different locations for frequencies ranging from $0.05$ to $25$. The results are reported and compared to the theory in figure \ref{systematique_cylindre}. When $\omega >2$, very good agreement is obtained between theory and numerics. In this regime, no inertial modes can be directly excited by the forcing. The hypothesis of Wang (i.e. no inertial mode forcing) is then satisfied. Note in particular that the zonal flow in this limit can be prograde. Good agreement is also observed for the smallest frequency ($\omega=0.05$) that we have considered. For this frequency, we suspect that no inertial mode is excited due to the finite length of the cylinder and the relatively large value of our Ekman number. By contrast, for the other frequencies between $0$ and $2$, inertial modes are present in the system which lead to differences between theory and numerical results (see figure \ref{systematique_cylindre}(b)). This regime in which inertial modes are excited is studied in the next section.

%%%%%%%%%%%%%%%%%%%%%%%%%%%%%%%%%%%%%%%%%%
\subsection{Zonal flow in the presence of inertial modes}

{{ 
In this section, we shall try to provide some generic features of the zonal flow in the presence of inertial modes. Again, we consider small amplitude configurations where the boundary layer remains centrifugally stable. }}

\begin{figure}[h!]
\begin{center}
\begin{center}\includegraphics{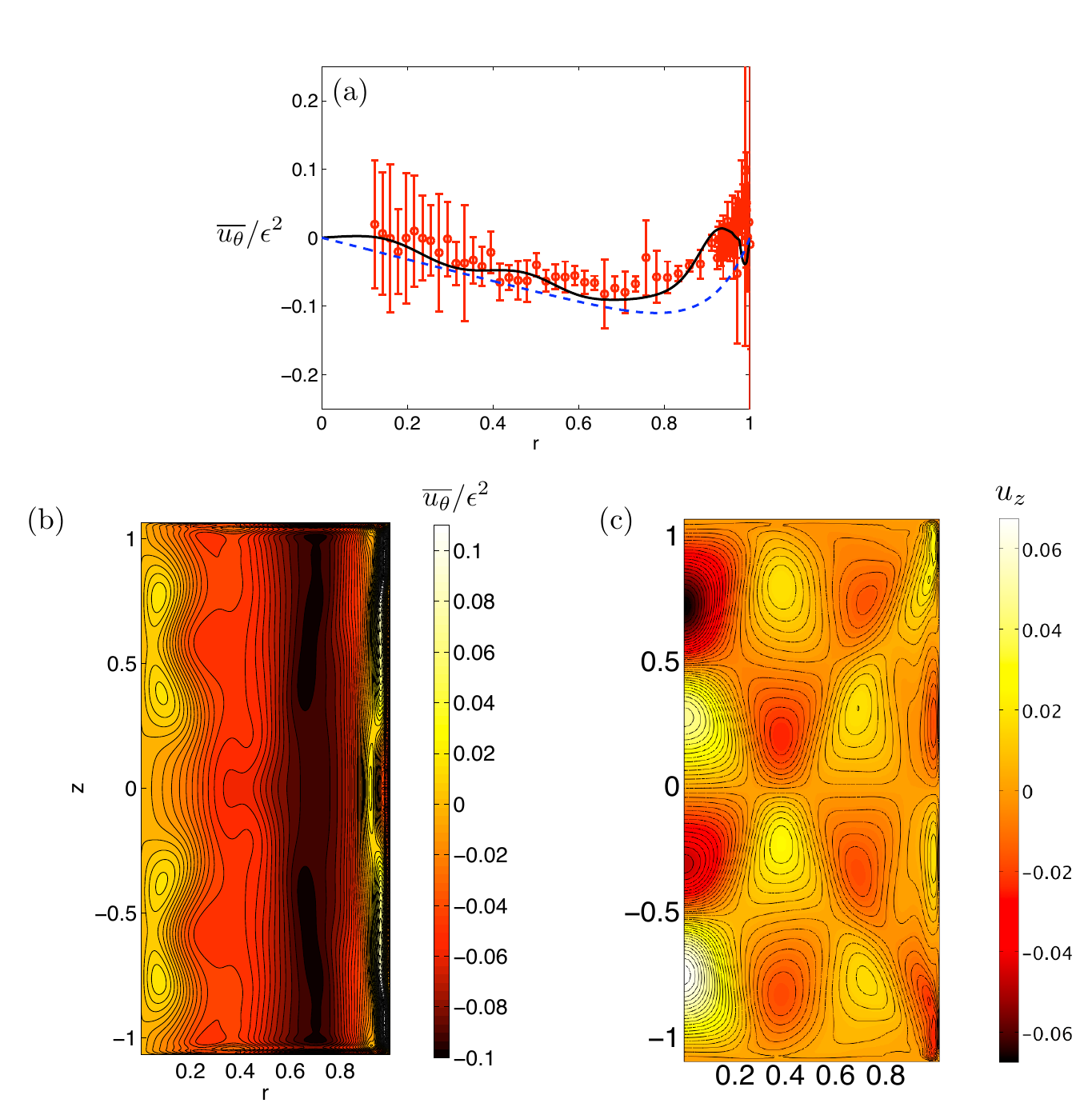}\end{center}
\caption{Comparison between experimental, theoretical and numerical results in a cylinder of aspect ratio $\alpha=1.0686$  for $\epsilon=0.4$, $E=10^{-4}$, $\omega=1$. (a) Mean zonal flow $\overline{u_\theta}/\epsilon^2$ at $z=0.22$. Red points together with the red error bars are the experimental measurements of Noir et al.\cite{noir2010} using LDV technics, black solid line and blue dashed line are the numerical results and the  theoretical prediction, respectively. (b) Contour 
of $\overline{u_\theta}/\epsilon^2$ in the $(r,z)$ plane. The mean zonal flow is obtained by averaging the azimuthal velocity over three libration periods.  (c) Instantaneous axial velocity contours (at time $t=2\pi/\omega$).}
\label{comparison_profile_noir}
\end{center}
\end{figure}

{{  Experimental results have been obtained by Noir et al.\cite{noir2010}. One of their configurations corresponds to a situation with a stable boundary layer. Their mean zonal flow measurements have been made at 
a fixed location $z=0.22$ in a cylinder of aspect ratio $\alpha =1.0686$ for the parameters $\om=1$, $\epsilon=0.4$, $E=10^{-4}$. Their results are reported in figure \ref{comparison_profile_noir}(a) together with our computational
results and Wang's prediction.  }}

{{  One can see that numerical and experimental results are in good agreement but depart from Wang's curve. This is not surprising as for $\om=1$  and $\alpha=1.0686$, an inertial mode 
is resonantly excited whose structure can be seen in figure \ref{comparison_profile_noir}(c). In this figure, we can indeed observe a well-defined pattern in the bulk of the fluid which
corresponds to the structure of the resonant inertial mode. The presence of this mode modifies the spatial structure of the mean zonal flow which now strongly depends on the axial coordinate, as illustrated
in figure \ref{comparison_profile_noir}(b). }}

\begin{center}
\begin{figure}[h]
\begin{center}\includegraphics{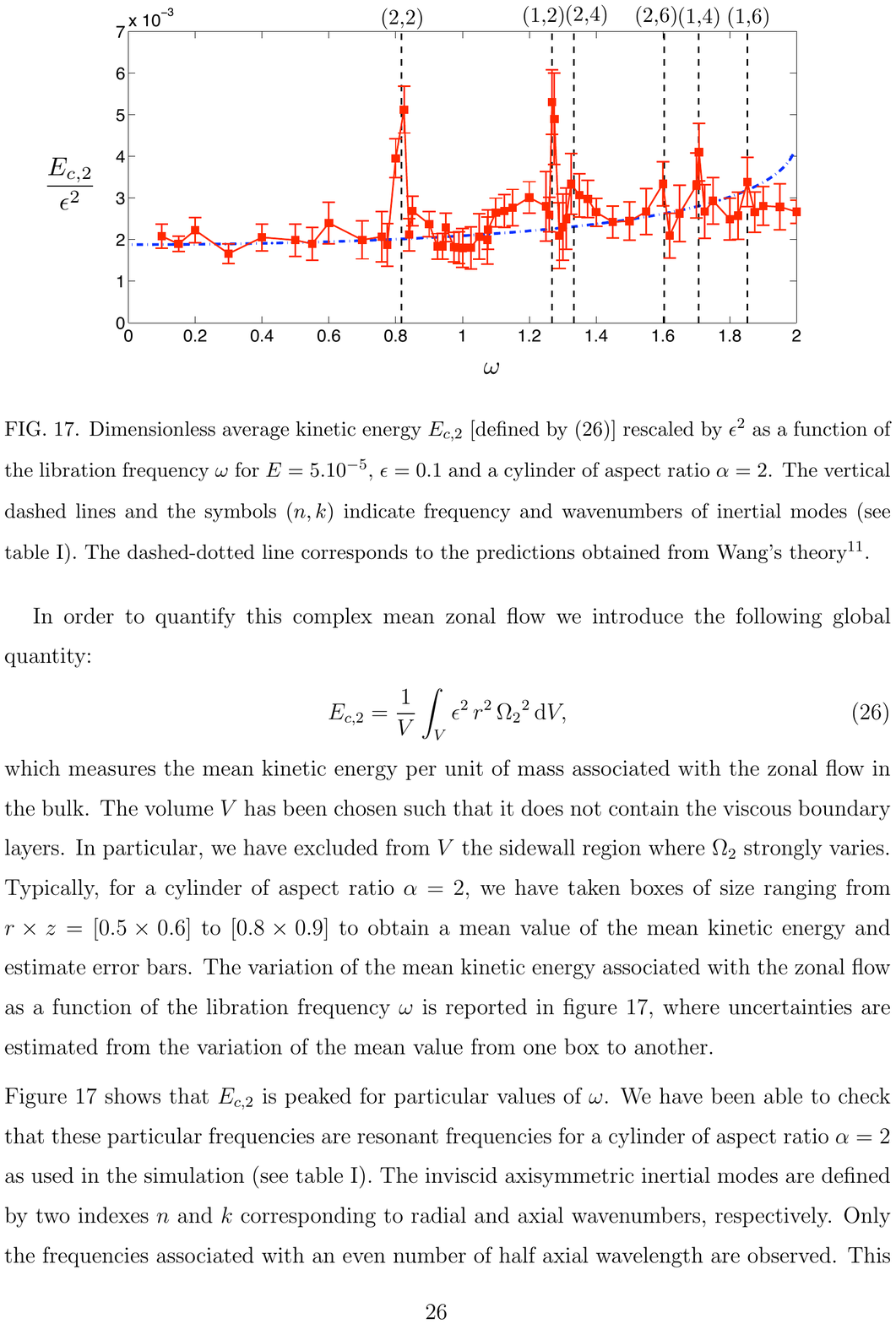}\end{center}
\caption{Dimensionless average kinetic energy $E_{c,2}$ [defined by (\ref{defEc2})] rescaled by $\epsilon^2$ as a function of the libration frequency $\omega$ for $E=5.10^{-5}$, $\epsilon=0.1$ and a cylinder of aspect ratio $\alpha=2$. The vertical dashed lines and the symbols $(n,k)$ indicate frequency and wavenumbers of inertial modes (see table I). The dashed-dotted line corresponds to the predictions obtained from Wang's theory\cite{wang1970}.}\label{syst_inertiel2}
\end{figure}
\end{center}

{{
In order to quantify this complex mean zonal flow we introduce the  following global quantity:
\begin{eqnarray}\label{defEc2}
E_{c,2}  & = & \frac{1}{V}\int_V\epsilon^2\,r^2\,{\Omega_2}^2 \,\text{d} V,
\end{eqnarray}
\noindent which measures the  mean kinetic energy per unit of mass associated with the zonal flow in the bulk. The volume $V$ has been chosen such that it does not contain the 
viscous boundary layers. In particular, we have excluded from $V$ the sidewall region  where $\Omega_2$  strongly varies.  }}
Typically, for a cylinder of aspect ratio $\alpha=2$, we have taken boxes of size ranging from $r\times z=[0.5 \times 0.6]$ to $[0.8 \times 0.9]$ {{to obtain a mean value of the mean kinetic energy and estimate error bars}}. The variation of the mean kinetic energy associated with the zonal flow as a function of the libration frequency $\omega$ is reported in figure \ref{syst_inertiel2}, where uncertainties are estimated from the variation of the mean value from one box to another.

 Figure \ref{syst_inertiel2} shows that $E_{c,2}$ is peaked for particular values of $\omega$. We have been able to check that these particular frequencies are resonant frequencies for a cylinder of aspect ratio $\alpha=2$ as used in the simulation (see table I). {{The inviscid axisymmetric inertial modes are defined by two indexes $n$ and $k$ corresponding to  radial and axial wavenumbers, respectively.}} 
 Only the frequencies associated with an even number of half axial wavelength are observed. This is in agreement with the constraint imposed by the forcing. Indeed, libration induces an azimuthal flow which is even with respect to the mid-plane ($z=0$). Modes with an odd number of half-axial wavelength have an odd azimuthal velocity component and therefore they cannot be forced directly. We can also notice that the peaks are the largest for the frequencies associated with the smallest wavenumbers and that frequencies corresponding to $n \geq 3$ are not visible. This is a viscous effect which tends to damp the largest wavenumber modes. Note finally that the general trend of $E_{c,2}$ with respect to $\omega$ still follows the theory by Wang\cite{wang1970}.

\begin{center}
\begin{figure}[h!]
\begin{center}
\includegraphics{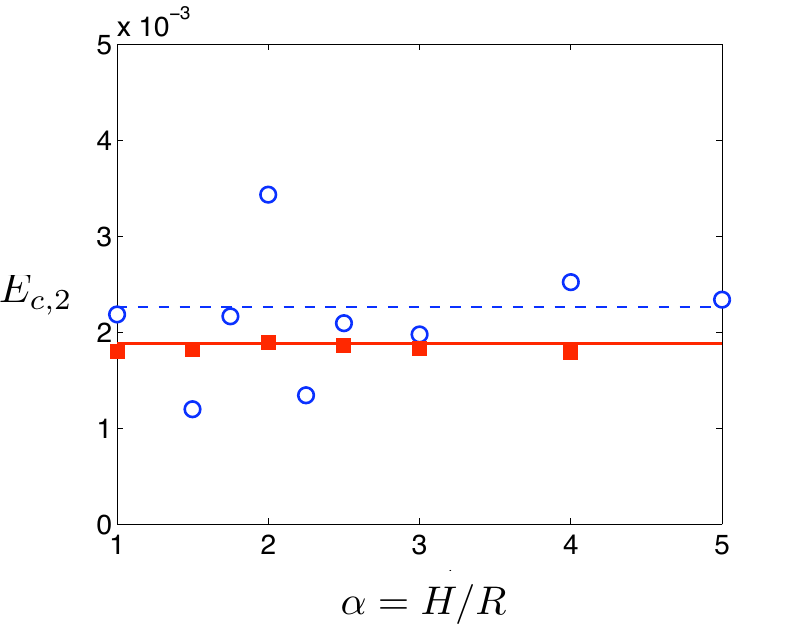}
\end{center}
\caption{Kinetic energy of the mean zonal flow for $\omega=0.1$ (red squares) and $\omega=1.27$ (blue circles) for various values of the aspect ratio of the cylinder $\alpha=R/H$ and $E=5.10^{-5}$, $\epsilon=0.1$. The red continuous line and the blue dashed line are respectively the theoretical value of the kinetic energy in the absence of inertial modes for $\omega=0.1$ ($E_{c,2} \simeq 1.88 \times 10^{-3}$) and $\omega=1.27$  ($E_{c,2} \simeq 2.23 \times 10^{-3}$). The frequency $\omega=1.27$ corresponds to the frequency of the mode $(1,2)$ for a cylinder of aspect ratio $\alpha=2$ (see table I).}\label{syst_hauteur}
\end{figure}
\end{center}

\begin{table}
\label{table:inertial}
 \begin{tabular}{@{}l|c|c|c|c|c|c}
& $n=1$ & $n=2$ & $n=3$  \\
$k=1$ & $0.7586$ & $0.4370$ & $0.3052$  \\
$k=2$ & $1.2681$ & $0.8174$ &  $0.5901$   \\
$k=3$ & $1.5518$   & $1.1152$  &  $0.8406$  \\
$k=4$ & $1.7075$  & $1.3343$ &   $1.0509$   \\
$k=5$ & $1.7975$  &  $1.4916$  &  $1.2222$    \\
$k=6$ & $1.8527$ &   $1.6043$   & $1.3592$   \\
$k=7$ & $1.8886$ &   $1.6860$   & $1.4680$   \\
$k=8$ & $1.9130$ &   $1.7463$   & $1.5544$   \\
\end{tabular}
\caption{Inviscid frequency of axisymmetric inertial modes as a function of their radial wavenumber $n$ and axial wavenumber $k$ for a cylinder of aspect ratio $\alpha=2$ (see Greenspan\cite{greenspanbook}).}
\end{table}  
 Resonances can also be seen when the aspect ratio of the cylinder is varied for a fixed frequency. Figure \ref{syst_hauteur} shows the variation of the zonal flow kinetic energy as a function of the aspect ratio for two different frequencies: $\omega=0.1$ and $\omega=1.27$. 
 For $\omega=0.1$, $E_{c,2}$ remains constant and equal to the prediction by Wang\cite{wang1970} as no inertial mode is excited for the parameters which have been considered. 
 For $\omega=1.27$, a clear peak is observed at $\alpha=2$, in agreement with the excitation of the inertial mode of wavenumbers $(n,k)=(1,2)$ (see table I). The excitation of this mode is also responsible for the deviation from Wang's theory at $\alpha=4$. It is interesting to note that resonance could also decrease the zonal flow kinetic energy as observed for $\alpha=1.5$ and $\alpha=2.2$ {{where modes $(3,4)$ and $(2,2)$ are excited respectively.}}

\begin{center}
\begin{figure}[h]
\begin{center}\includegraphics{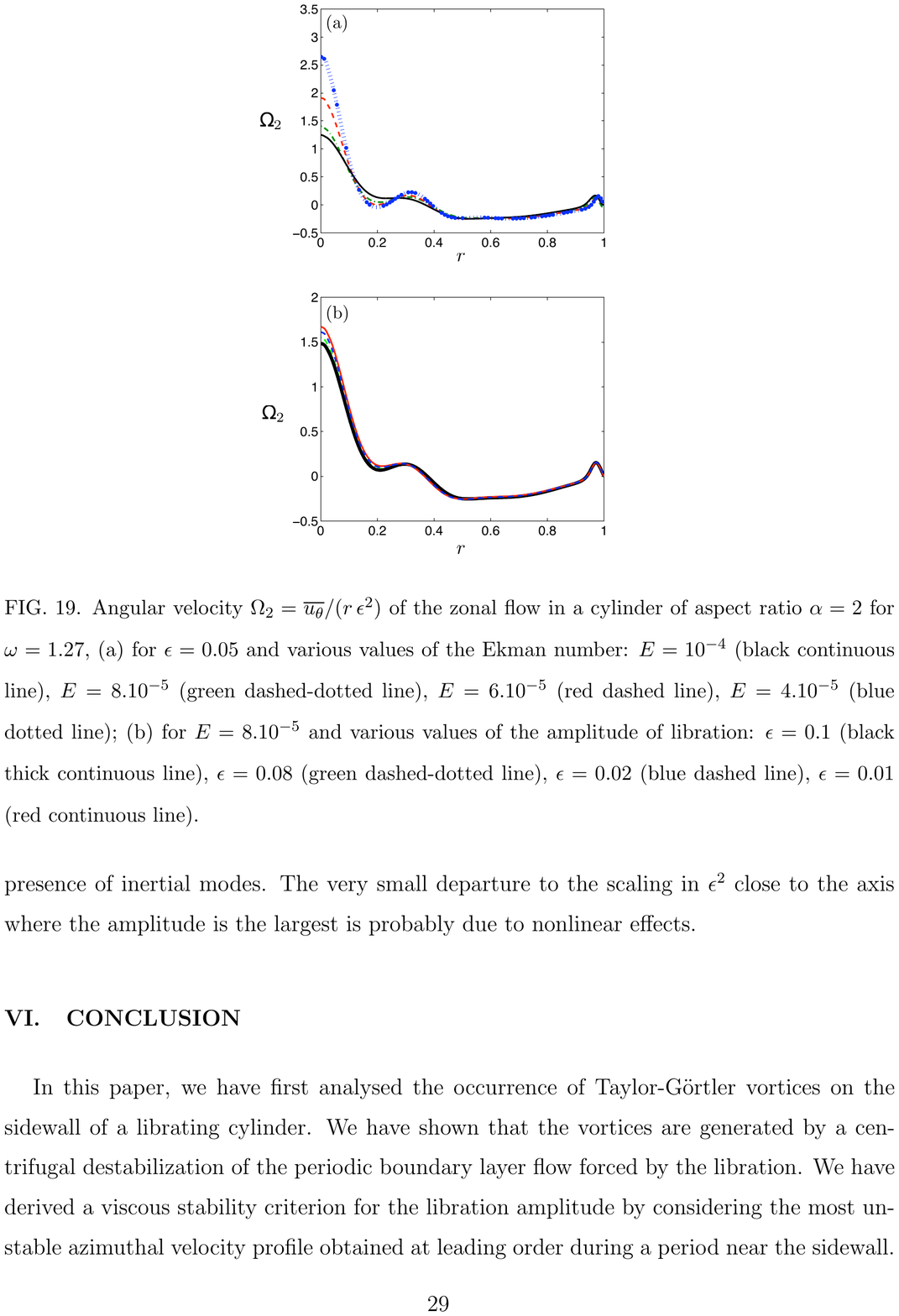}\end{center}
\caption{Angular velocity $\Omega_2=\overline{u_\theta}/(r\,\epsilon^2)$ of the zonal flow in a cylinder of aspect ratio $\alpha=2$ for $\omega=1.27$, (a) for $\epsilon=0.05$ and various values of the Ekman number: $E=10^{-4}$ (black continuous line), $E=8.10^{-5}$ (green dashed-dotted line), $E=6.10^{-5}$ (red dashed line), $E=4.10^{-5}$ (blue dotted line); (b) for $E=8.10^{-5}$ and various values of the amplitude of libration: $\epsilon=0.1$ (black thick continuous line), $\epsilon=0.08$ (green dashed-dotted line), $\epsilon=0.02$ (blue dashed line), $\epsilon=0.01$ (red continuous line).}\label{syst_inertielE}
\end{figure}
\end{center}

We have seen that, when an inertial mode is excited, the zonal flow is no longer homogeneous in the bulk [see figure \ref{comparison_profile_noir}(b)]. The radial structure of the angular velocity $\Omega_2$ of the zonal flow at a fixed axial location is illustrated in figure \ref{syst_inertielE} when the mode $(n,k)=(2,1)$ is excited. In this figure, the effects of the variations of the Ekman number (figure \ref{syst_inertielE}(a)) and of the libration amplitude (figure \ref{syst_inertielE}(b)) are analysed. A clear dependence with respect to the Ekman number can be noted near the axis. This dependence is not associated with a viscous detuning of the inertial mode frequency as the viscous correction is negligible for the Ekman number we have considered\cite{wedemeyer1964}. Instead, we think that it could be due to strong nonlinear interactions occurring in the region close to the axis where the viscous shear layers emitted from the corners encounter.

{{ As the strength of the internal shear layers depends on the Ekman number, it is not surprising to observe an increase of the zonal flow when the Ekman number decreases. Unfortunately, our range of accessible Ekman numbers did
not allow us to provide a scaling for the zonal flow. Note that when the zonal flow is generated by tidal forcing, a scaling in $E^{-3/10}$ was observed experimentally in a sphere\cite{morize2010}, and a scaling in $E^{-3/2}$ was suggested numerically in a spherical shell\cite{tilgner2007}.}}
 
 In figure \ref{syst_inertielE}(b), we observe that the zonal flow remains proportional to $\epsilon^2$ even in the presence of inertial modes. The very small departure to the scaling in $\epsilon^2$ close to the axis where the amplitude is the largest is probably due to nonlinear effects. 
 
%%%%%%%%%%%%%%%%%%%%%%%%%%%%%%%%%%%%%%%%%%%
%%%%%%%%%%%%%%%%%%  Conclusion %%%%%%%%%%%%%%%%%%%
%%%%%%%%%%%%%%%%%%%%%%%%%%%%%%%%%%%%%%%%%%%
  
  \section{Conclusion}

In this paper, we have first analysed the occurrence of Taylor-G\"ortler vortices on the sidewall of a librating cylinder. We have shown {{that the vortices are generated by a centrifugal destabilization}} of the periodic boundary layer flow forced by the libration. {{We have derived a viscous stability criterion for the libration amplitude by considering the most unstable azimuthal velocity profile obtained at leading order during a period near the sidewall. This criterion for the threshold amplitude provides a scaling in $E^{1/2}$ and a dependency with respect to $\omega$ in qualitative good agreement with the numerical and experimental data. Quantitative differences have been observed and some explanations have been proposed. As the instability criterion only depends on the local structure of the boundary layer, which just has to be vertical and axisymmetric, it can be applied to other configurations. In particular, we could apply the criterion to a librating sphere in the boundary layer close to the equator.}} The nonlinear regime of the instability has also been studied numerically. We have shown that the Taylor-G\"ortler vortices can force coherent structures in the bulk of the cylinder whose characteristics exhibit a preferred orientation around $45\!\char23$ independent of the libration frequency.

Below the instability threshold, we have considered the mean zonal flow generated by libration. We have shown that the theory by Wang\cite{wang1970} captures the main characteristics of the zonal flow in the bulk for small frequencies and for frequencies larger than $2$, i.e. when inertial modes are not directly excited.
However, an unexplained discrepancy has been systematically observed close to the sidewall. The zonal flow in the presence of excited inertial modes has also been studied numerically. We have shown that the peaks of the zonal flow kinetic energy as the libration frequency or the cylinder aspect ratio is varied correspond to the resonant excitation of large-scale inertial modes. The dependence of the zonal flow with respect to the Ekman number and the libration amplitude has also been documented.

{{ All the numerical simulations have been carried out with an axisymmetric code. Despite this constraint, we have seen that, in a librating cylinder, a mechanism of generation of inertial waves, different from direct forcing and period doubling, could be active when the sidewall boundary layer becomes turbulent. It would now be interesting to determine whether this process is active in 3D and how it depends on the geometry. In particular, a precise description of the dynamics of librating ellipsoidal containers \cite{chan2011,zhang2011} and the influence of the topography on the resulting flow would be of great interest.
}}

%%%%%%%%%%%%%%%%%%%%%%%%%%%%%%%%%%%%%%%%%%%
%%%%%%%%%%%%%%%%  Acknowledgement %%%%%%%%%%%%%%%%
%%%%%%%%%%%%%%%%%%%%%%%%%%%%%%%%%%%%%%%%%%%
  
  \section*{Acknowledgements}
  
We thank J\'er\^ome Noir for fruitful discussions and for having provided the experimental data of figures \ref{noir} and \ref{comparison_profile_noir}(a).  %SLD

%%%%%%%%%%%%%%%%%%%%%%%%%%%%%%%%%%%%%%%%%%%
%%%%%%%%%%%%%%%%%%  Appendix %%%%%%%%%%%%%%%%%%%
%%%%%%%%%%%%%%%%%%%%%%%%%%%%%%%%%%%%%%%%%%%

  \section*{Appendix}
    
  {{The expression for the mean angular velocity, $\Omega_2$, appearing in formula (\ref{wangcomposite}) are given below for $\omega <2$ and $\omega >2$ respectively (Note that these expressions were first derived by Wang\cite{wang1970} using a different notation).}}
\begin{widetext}
For $\omega<2$:
\begin{eqnarray}
\Omega_2  & = & \frac{(\omega+1)\,\sqrt{4+2\,\omega}+2}{8\,\sqrt{4+2\,\omega}\,(\omega^2+4\,\omega+5)}-\frac{(\omega-1)\,\sqrt{4-2\,\omega}-2}{8\,\sqrt{4-2\,\omega}\,(\omega^2-4\,\omega+5)}-\frac{1}{\sqrt{4-2\,\omega}\,\sqrt{4+2\,\omega}} \nonumber \\
& & +\frac{\sqrt{4-2\,\omega}\,\sqrt{4+2\,\omega}+\omega(\sqrt{4+2\,\omega}-\sqrt{4-2\,\omega})-4}{\sqrt{4-2\,\omega}\,\sqrt{4+2\,\omega}\,(2+\sqrt{4+2\,\omega}\,\sqrt{4-2\,\omega})(\sqrt{4+2\,\omega}+\sqrt{4-2\,\omega})},
\label{wangresult1}
\end{eqnarray}

and for $\omega >2$
\begin{eqnarray}
\Omega_2  & = & \frac{(\omega+1)\,\sqrt{2\,\omega+4}+2}{8\,\sqrt{2\,\omega+4}\,(\omega^2+4\,\omega+5)}-\frac{(\omega-3)\,\sqrt{2\,\omega-4}+2}{8\,\sqrt{2\,\omega-4}\,(\omega^2-4\,\omega+5)} \nonumber \\
& &+\frac{2\,\omega-\sqrt{2\,\omega+4}\,\sqrt{2\,\omega-4}}{2\,\omega\,\sqrt{2\,\omega+4}\,\sqrt{2\,\omega-4}}+\frac{2\,\omega\,(\omega^2-6)-\sqrt{2\,\omega+4}\,(\omega^2+3\,\omega-12)}{4\,\omega\,(\omega^2-3)\,\sqrt{2\,\omega+4}\,\sqrt{2\,\omega-4}} \nonumber \\
& & -\frac{\sqrt{2\,\omega-4}\,(\omega^2-3\,\omega-12)+3\,\omega\,\sqrt{2\,\omega+4}\,\sqrt{2\,\omega-4}}{4\,\omega\,(\omega^2-3)\,\sqrt{2\,\omega+4}\,\sqrt{2\,\omega-4}}.
\label{wangresult2}
\end{eqnarray}
\end{widetext}

%%%%%%%%%%%%%%%%%%%%%%%%%%%%%
%%%%%%%%%%% Bibliography %%%%%%%%%%%
%%%%%%%%%%%%%%%%%%%%%%%%%%%%%

%

\end{document}